\documentclass[lettersize,journal]{IEEEtran}
\usepackage{amsmath,amsfonts}
\usepackage{amssymb}
\usepackage{algorithmic}
\usepackage[caption=false,font=footnotesize,labelfont=sf,textfont=sf]{subfig}
\usepackage{stfloats}
\usepackage{url}
\usepackage{graphicx}
\usepackage{cite}
\usepackage{float}
\usepackage{subfig}
\usepackage{multirow}
\usepackage{color}
\usepackage{makecell}
\usepackage{booktabs}
\usepackage[ruled,linesnumbered]{algorithm2e}
\usepackage[colorlinks,citecolor=red,linkcolor=blue]{hyperref}

\newtheorem{theorem}{Theorem}

\newtheorem{lemma}{Lemma}
\newtheorem{property}{Property}
\makeatletter

\newcommand{\Rmnum}[1]{\expandafter\@slowromancap\romannumeral #1@}
\makeatother

\def\BibTeX{{\rm B\kern-.05em{\sc i\kern-.025em b}\kern-.08em
    T\kern-.1667em\lower.7ex\hbox{E}\kern-.125emX}}
\usepackage{balance}
\begin{document}

\title{{A Statistical Model of Bursty Mixed Gaussian-impulsive Noise: Model and Parameter Estimation}}

\author{Tianfu Qi, \emph{Graduate Student Member, IEEE}, Jun Wang, \emph{Senior Member, IEEE},
\thanks{Tianfu Qi, Jun Wang are with the National Key Laboratory of Wireless Communications, University of Electronic Science and Technology of China, Chengdu 611731, China (e-mail: 202311220634@uestc.edu.cn)}
}

\markboth{Qi \MakeLowercase{\textit{et al.}}: A Statistical Model of Bursty Mixed Gaussian-impulsive Noise and Its Applications}%
{Shell \MakeLowercase{\textit{et al.}}: A Sample Article Using IEEEtran.cls for IEEE Journals}

\maketitle

\begin{abstract}
Non-Gaussian impulsive noise (IN) with memory exists in many practical applications. When it is mixed with white Gaussian noise (WGN), the resultant mixed noise will be bursty. The performance of communication systems will degrade significantly under bursty mixed noise if the bursty characteristic is ignored. A proper model for the bursty mixed noise and corresponding algorithms needs to be designed to obtain desirable performance but there is no such model reported to the best of our knowledge. The important problem is addressed in the two-part paper. In the first part, we propose a closed-form heavy-tailed multivariate probability density function (PDF) that to model the bursty mixed noise. This model is the weighted addition of gaussian distribution and student distribution. Then, we present the parameter estimation method based on the empirical characteristic function of the proposed model and analyze the performance of the parameter estimation. Numerical results show that our proposed bursty mixed noise model matches the measured bursty noise well. Meanwhile, the parameters of the proposed noise model can be accurately estimated in terms of mean square error (MSE).
\end{abstract}


\section{Introduction}
Different from the white Gaussian noise (WGN), there are many outliers in the non-Gaussian impulsive noise (IN) with quite short duration but large power, which lead to significant performance deterioration of communication systems \cite{paper8}. IN exists in many practical scenarios including lightning \cite{paper4}, IoT wireless systems \cite{paper5,paper6}, wide-band powerline systems \cite{paper3}, underwater acoustic communication \cite{paper2}, etc.

\subsection{Related literature}
Several statistical models for IN have been proposed. For example, Middleton proposed the Middleton class A/B/C (MCA/B/C) models for IN with respect to different scenarios. Middleton's models can match practical noise well, but the corresponding probability density functions (PDF) are quite complicated \cite{paper7}. Nikias proved that the MCB model is equivalent to the symmetric $\alpha$ stable (S$\alpha$S) distribution, which has a concise characteristic function (CF). Nowadays, the S$\alpha$S distribution has been widely used for modeling and analyzing the IN \cite{paper8}. Unfortunately, S$\alpha$S distribution with heavy tail does not admit a close-form PDF expression. Therefore, it is very challenging to design optimal signal processing algorithms based on the S$\alpha$S distribution.

The above mentioned IN models belong to the category of white IN model by assuming that the noise samples are mutually independent. However, this assumption is not always satisfied in practice. Different from the white IN model, colored IN models consider the correlation among noise samples. For example, Mahmood finds that snapping shrimp behaviors generate impulsive noise clusters with high peak-to-peak levels \cite{paper9}. In \cite{paper4}, the measured atmospheric radio noise in the range 10 Hz to 60 kHz also has the memory characteristic and the outliers preserve a longer duration. To describe the property, Chrissan et al. utilize the Poisson cluster model to analyze the features of sferic noise such as the cluster interarrival times, amplitude distribution and IN cluster length \cite{paper4}. Simulation results show this model fits measured atmospheric noise very well. However, the adopted model has no close-formed PDF of amplitude distribution. Meanwhile, the model is composed of several distributions with many parameters, which makes it quite inconvenient to generate noise samples and design optimal signal detection algorithms.

Another kind of statistical models are proposed based on the Markov chain \cite{book1} to describe the noise state transitions. For example, Markov-Gaussian (MG) and Markov-Middleton (MM) models leverage the mixture Gaussian distribution to describe the colored IN \cite{paper10,paper11}. These models have a unique state transition matrix associated with the noise PDF. In each state, the noise amplitude is assumed to follow Gaussian distribution with various variances and the state with larger variance represents IN \cite{paper10}. They have analytical PDF but not heavy-tailed, which makes the generated noise samples are far from the real bursty IN. Some noise models based on modified Markov chains such as the partitioned Markov chain and semi-hidden Markov chain are also proposed \cite{paper12,paper13}. However, there is no essential difference compared with the MM and MG.

The previous mentioned bursty IN models could not admit thick-tailed properties and describe the correlation between arbitrary noise samples. A. Mahmood and M. Chitre discover that the scatter plots of two neighboring time samples follow the elliptical distribution, and apply the $\alpha$-sub Gaussian ($\alpha$SG) distribution to model the IN with memory \cite{paper9}. $\alpha$SG model is represented by a random vector, for which each element follows the same S$\alpha$S distribution. The distribution is not isotropic and therefore, it can match the measured underwater noise better compared with models based on the mixture Gaussian distribution.

As mentioned before, the S$\alpha$S distributed noise has no close-formed PDF except for Gaussian and Cauchy cases \cite{book2}. Meanwhile, the background WGN is not considered in this model. Thus, the generated noise samples based on the $\alpha$SG distribution also cannot well match real bursty noise. In fact, the WGN is inevitable due to the Brownian motion of electrons. Unfortunately, it is difficult to obtain the analytical model for the mixed noise of IN and WGN due to the complexity of IN. Sureka et al. propose a white mixed noise model with closed-form based on the property of S$\alpha$S distribution \cite{paper14}. The optimal demodulation performance of MSK signals based on this mixed noise model is further verified in \cite{paper20}. However, to the best of our knowledge, there is no analytical model reported to describe the mixed noise of IN and WGN with memory, i.e., bursty mixed Gaussian-impulsive noise.

\subsection{Motivation and Contributions}
Obviously, it will introduce significant performance degradation if the detection algorithm for white IN is directly applied under bursty mixed noise \cite{paper10}. Therefore, it is of vital significance to obtain a concise and analytical model of bursty mixed Gaussian-impulsive noise to design optimal signal detection algorithms. In this part of the two-part paper, we address this issue by building a statistical model of bursty mixed Gaussian-impulsive noise, and explore the method to estimate the parameters. Our contributions are summarized as follows,
\begin{enumerate}
\item{First, we propose a new close-formed multivariate statistical model for bursty mixed noise based on Gaussian distribution and student distribution. This proposed Gaussian-student (GS) model is an elliptic and heavy-tailed distribution. Meanwhile, it is a general model that can degrade to WGN and pure bursty IN with specific parameters.}
\item{Second, we present the parameter estimation algorithms of our proposed GS model based on the characteristic function (CF) and PDF. In addition, we analyze the performance of the proposed parameter estimation method in terms of unbiasedness, consistency, etc. Numerical results verify that the proposed estimation algorithms can achieve excellent performance in term of mean square error (MSE) under various scenarios.}
\item{Last but not least, the similarity between two high-dimension PDF is challenging to quantized by Kullback-Leibler (KL) divergence directly due to the unacceptable computation complexity. Therefore, we utilize three important physical characteristics of the bursty mixed noise including  to measure the similarity of different noise models. Simulation results show that this GS model can well describe the cluster and impulsive features of bursty mixed noise.}
\end{enumerate}

\subsection{Organization and Notations}
The remainder of the paper is organized as follows. In section \ref{statisticalmodel}, the new statistical model is proposed and its properties are discussed. In section \ref{parameterestimation}, the parameter estimation algorithm is derived and the estimation performance is analyzed. Simulation results are presented in section \ref{simulation}. Finally, We conclude this paper in section \ref{conclusion}.

$\mathbf{Notations}$: We denote the random vector and its realization by bolded uppercase and lowercase letters such as $\mathbf{X}$ and $\mathbf{x}$, respectively. Unless stated otherwise, the vector is column vector in this paper, i.e., $\mathbf{x}=[x(1),x(2),\cdots,x(p)]^\top$. Specifically, we denote the $i$-th sample vector as  $\mathbf{x}_i=[x_i(1),x_i(2),\cdots,x_i(p)]^\top$. The notations `$\hat{\Sigma}$' and `$\Sigma^\top$' represent the estimate and transpose of matrix $\Sigma$, respectively. We use the notations `$\Sigma_{i,j}$' and `$\Sigma(i,j)$' to denote the block matrix and the element of the $i$-th row and $j$-th column of $\Sigma$, respectively. For instance, the matrix $\Sigma$ is partitioned as follows:

\begin{equation}
\Sigma=
\left[
\begin{array}{cc}
  \mathbf{A} & \mathbf{B} \\
  \mathbf{C} & \mathbf{D}
\end{array}
\right],
\end{equation}

Then, we denote $\mathbf{C}$ by $\Sigma_{2,1}$ and $\mathbf{B}$ by $\Sigma_{1,2}$. The random variable (RV) and its samples are separately presented by unbolded uppercase and lowercase letters, i.e., $X$ and $x$. We utilize the notation `$P(\cdot)$' to represent the probability of a set. The covariance between RVs $X$ and $Y$ is denoted by $Cov(X,Y)$ and $\mathbb{E}_{\mu_{X}}(X)$ is the expectation of RV $X$ where $\mu_{X}$ is the probability measure of $X$. The notation `$\Vert\cdot\Vert$' denotes the Frobenius norm. The element number of a set is denoted as $\mathbf{card}(\cdot)$. $\mathbb{R}^{p}$, $\mathbb{Z}^{+}$ and $\mathbb{S}^{p}_{++}$ separately represent the $p$-dimensional domain of real numbers, positive integer set and the set of $p\times p$ positive definite matrix. $\exp(\cdot)$ is the exponential function and $\det(\cdot)$ is the determinant operation. $\mathbf{e}_p$ denotes the $p$-dimensional all-one vector. Finally, we denote the $p$-order identical matrix by $\mathbf{I}_p$ and the Gamma function to be $\Gamma\left(a\right)=\int_{0}^{+\infty}t^{a-1}\exp(-t)dt$.

\section{Noise Model}\label{statisticalmodel}
In this section, we first concisely describe $\alpha$SG model and its properties used in the following derivation. Then, we propose a new statistical model of bursty mixed noise based on Gaussian and student distributions. Finally, some useful properties and the generality of the proposed model are discussed.

\subsection{$\alpha$SG model of bursty IN}
$\alpha$ stable distribution is a class of distribution with a heavy tail except the special case of Gaussian distribution \cite{book2}. The RV $S$ following $\alpha$ stable distribution can be expressed as $S\sim \mathcal{S}(\alpha,\beta,\gamma,\mu)$. $\alpha$ is the characteristic parameter representing the thickness of the tail. $\gamma$ is the scale parameter that describes the power of the RV. Specifically, we have $2\gamma^2=\sigma^2$ for normal distribution with variance $\sigma^2$. $\beta$ and $\mu$ are skewness and position parameters, respectively. $S$ becomes S$\alpha$S distribution when $\beta=\mu=0$, which is mainly considered in this paper. Note that $\alpha$ stable distribution does not admit analytic PDF except Gaussian distribution ($\alpha=2$) and Cauchy distribution ($\alpha=1$).

Different from white IN, bursty IN admits the behavior of cluster \cite{paper4}. Based on the S$\alpha$S distribution, $\alpha$SG model has been used to describe the IN with memory \cite{paper9}. Define the following random vector
\begin{equation}
\mathbf{X}=[X(1),X(2),\cdots,X(p)]^\top,p\in \mathbb{Z}^{+},
\end{equation}
where the $X(i),i=1,\cdots,p$ are identical but not independent RVs and $p$ is the memory order. The $\alpha$SG distribution can be expressed by the product of independent $\alpha$ stable RV and Gaussian random vector, i.e.,
\begin{equation}
\underbrace{[X(1),X(2),\cdots,X(p)]^\top}_{\mathbf{X}}=\sqrt{S}\underbrace{[G(1),G(2),\cdots,G(p)]^\top}_{\mathbf{G}},
\end{equation}
where the $G(i),i=1,\cdots,p$ are identically distributed normal RVs with covariance matrix $\Sigma\in \mathbb{S}^{p}_{++}$. The $S$ is a $\alpha$ stable RV and $S\sim \mathcal{S}(\alpha/2,1,2\left[\cos(\alpha\pi/4)\right]^{2/\alpha},0)$ \cite{book2}. In \cite{paper15}, the PDF of $\mathbf{X}$ is investigated based on the radial RV and variable transformation to get
\begin{equation}\label{asymptotic_order}
\lim_{r\rightarrow+\infty} r^{\alpha+p}f_{\alpha}(\mathbf{x})=c,
\end{equation}
where $r=\Vert \Sigma^{-1/2}\mathbf{x}\Vert$ and $\Sigma^{1/2}$ is the Cholesky decomposition of $\Sigma$, i.e., $\Sigma=(\Sigma^{1/2})^\top\Sigma^{1/2}$. Here, $c$ is a constant and $f_{\alpha}(\mathbf{x})$ represents the distribution of $\mathbf{X}$.

Based on \eqref{asymptotic_order}, we conclude that the distribution with tail decay order $(\alpha+p)$ can be used to equivalently describe the bursty IN with memory order $p$. We adopt the multivariate student distribution for two reasons. First, the degree of freedom (DOF) of student distribution can be flexibly adjusted to emulate the heavy-tailed characteristic. Then, the student distribution is general since it degrades to Cauchy distribution with DOF to be 1, which coincides with $\alpha$SG distribution and becomes Gaussian distribution when DOF is infinite.

\subsection{Multivariate statistical model for bursty mixed noise}
We consider that the bursty mixed noise consists of IN with memory and WGN. These two noise components are assumed to be mutually independent. This assumption is reasonable because WGN and IN are introduced by different sources.

The bursty mixed noise vector $\mathbf{N}$ can be expresses as follows,
\begin{equation}\label{bursty_mixed_noise}
\mathbf{N}=\mathbf{N}_G+\mathbf{N}_I,
\end{equation}
where $\mathbf{N}_G$ and $\mathbf{N}_I$ denote the WGN vector and IN vector with memory, respectively.  Without loss of generality, the correlation between noise samples is assumed to be weak and can be omitted if the corresponding time interval is larger than a threshold. That is the memory order of $\mathbf{N}$ is finite by following the Markov property. Then, we set $\mathbf{N}=[N(1),N(2),\cdots,N(p)]^\top$ where $p$ is the memory order.

The well-known multivariate Gaussian distribution with identity covariance matrix can be used to model the WGN $\mathbf{N}_G$ in \eqref{bursty_mixed_noise}. To model the bursty IN $\mathbf{N}_I$ in \eqref{bursty_mixed_noise}, we need a distribution with heavy tail to describe the outliers and the correlation between noise samples. Motivated by the feature given in \eqref{asymptotic_order}, we propose a weighted sum of multivariate Gaussian and student distributions to model the mixed bursty noise $\mathbf{N}$. In what follows, we call this model as GS model for brevity.

Then, the PDF of mixed noise with memory can be given as follows,
\begin{equation}\label{bursty_mixed_noise_model}
f_M(\mathbf{n})=\rho k_1\exp\left(-\frac{\Vert\mathbf{n}\Vert^2}{4\gamma_g^2}\right) +\frac{(1-\rho)k_2}{\left(1+\Vert\Sigma^{-1/2}\mathbf{n}\Vert^2/\alpha\right)^\frac{\alpha+p}{2}},
\end{equation}
where $k_1$ and $k_2$ are normalization factors for WGN and bursty IN components, respectively. They can be calculated to be
\begin{equation}\label{k_1}
k_1=\frac{1}{(2\sqrt{\pi}\gamma_g)^p},
\end{equation}
\begin{equation}\label{k_2}
k_2=\frac{\Gamma(\frac{\alpha+p}{2})}{\Gamma(\frac{\alpha}{2})(2\gamma_s^2\alpha\pi)^{p/2}\sqrt{\det(\tilde{\Sigma})}},
\end{equation}
where we denote $\Sigma=2\gamma_s^2\tilde{\Sigma}$ and the diagonal elements of regularized covariance matrix $\tilde{\Sigma}$ are 1. The key parameters in \eqref{bursty_mixed_noise_model} are $\alpha$, $\gamma_g$, $\gamma_s$, $\rho$, $\Sigma$ and $p$. Similar to S$\alpha$S distribution, $\alpha$ is the characteristic parameter to represent the heaviness of the PDF tail. Mathematically, $\alpha$ herein belongs to the region of $(0,+\infty)$. For practical communication applications, we mainly focus on $\alpha\in(0,2)$, which just corresponds to $\alpha$SG model. $\gamma_g\in(0,+\infty)$ and $\gamma_s\in(0,+\infty)$ are the scale parameters of WGN and bursty IN, respectively. $\rho\in[0,1]$ is the weight parameter to adjust the mainlobe and tail section of the PDF, which describes the ratio of WGN and bursty IN. The smaller $\rho$ is, the more prominent cluster feature and impulsity the mixed noise has. $\Sigma\in\mathbb{S}^{p}_{++}$ and $p\in\mathbb{Z}^{+}$ are the covariance matrix and memory order of IN, respectively. The Cholesky decomposition of $\Sigma^{-1}$ exists since it is a positive definite matrix. Therefore, we have $\Sigma^{-1}=(\Sigma^{-1/2})^\top\Sigma^{-1/2}$ and $\Sigma^{-1/2}$ is a lower triangular matrix.

Compared with existing models \cite{paper4,paper9,paper11}, our proposed GS model \eqref{bursty_mixed_noise_model} has the following features and advantages:
\begin{enumerate}
\item{The GS model considers both the background WGN and the correlation between the IN samples. It could be flexible to match practical noise environments via adjusting the weight factor $\rho$ and memory order $p$.}
\item{In contrast to $\alpha$SG distribution, the GS model has a concise and closed-form PDF. It will facilitate the design of the optimal signal detection algorithms.}
\end{enumerate}

\subsection{Properties of GS model}
In this subsection, we discuss the properties of our proposed GS model that will be utilized in the following.
\begin{property}\label{property_1}
$f_M(\mathbf{n})$ is finite and continuous in $\mathbb{R}^p$.
\end{property}

This property is straightforward since there is no zero in the denominator of \eqref{bursty_mixed_noise_model}. Therefore, our proposed GS model is finite and continuous in the domain, which ensures that it is integrable.

\begin{property}\label{property_2}
The conditional PDF of $f_M(\mathbf{n})$ is
\begin{align}\label{conditional_PDF}
f_M(n|\mathbf{n}_1)=&\frac{1}{f_M(\mathbf{n}_1)}\Bigg[\rho k_1\exp\left(-\frac{n^2+\Vert\mathbf{n}_1\Vert^2}{4\gamma_g^2}\right) \nonumber\\
&+\frac{(1-\rho)k_2}{\left(1+\left(\Vert\Sigma_{1,1}^{-1/2}\mathbf{n}_1\Vert^2+(n-\delta)^2/\sigma\right)/\alpha\right)^\frac{\alpha+p}{2}}\Bigg],
\end{align}
where
\begin{equation}
\delta=\mathbf{n}_1^\top\Sigma_{1,1}^{-1}\Sigma_{1,2},
\end{equation}
\begin{equation}
\sigma=\frac{\det(\Sigma)}{\det(\Sigma_{1,1})},
\end{equation}
and $\Sigma$ is partitioned to
\begin{equation}
\Sigma=
\left[
\begin{array}{cc}
  \Sigma_{1,1} & \Sigma_{1,2} \\
  \Sigma_{1,2}^\top & \Sigma_{2,2}
\end{array}
\right],\Sigma_{1,1}\in \mathbb{S}^{p-1}_{++}.
\end{equation}
\end{property}

\begin{IEEEproof}
The proof is relegated to Appendix \ref{appendix_property_2}.
\end{IEEEproof}

Note that the $\mathbf{n}_1\in\mathbb{R}^{(p-1)\times1}$ and $\mathbf{n}=[\mathbf{n}_1^\top,n]^\top$. The conditional PDF can be used to generate the noise samples based on the acceptance-rejection method. Then, the next property provides the CF of the GS distribution.

\begin{property}\label{property_3}
The CF of $f_M(\mathbf{n})$ in \eqref{bursty_mixed_noise_model} is as follows,
\begin{align}\label{CF_GS}
\Phi_M(\mathbf{t})=&\rho\exp\left(-\gamma_g^2\Vert\mathbf{t}\Vert^2\right)+\frac{(1-\rho)\Gamma((\alpha+p)/2)\sqrt{\alpha}}{\Gamma(\alpha/2)(\alpha\pi)^{p/2}}\nonumber\\
&\times\int_{\mathbb{R}^p}\cos(\sqrt{\nu}\mathbf{t}^\top\Sigma^{1/2}\mathbf{y})\left(1+\mathbf{y}^\top\mathbf{y}\right)^{-\frac{\alpha+p}{2}}d\mathbf{y}.
\end{align}
\end{property}
\begin{IEEEproof}
The proof is relegated to Appendix \ref{appendix_property_3}.
\end{IEEEproof}

Even though there is no closed-form expression for \eqref{CF_GS}, it will still be leveraged for parameter estimation since the isotropic and anisotropic part are decoupled. In the next, we describe the generality of the GS model.
\begin{property}\label{property_4}
The GS model degenerates to special cases with parameters listed in Table \ref{specialcase_table}.
\begin{table}[htbp]
\begin{center}
\caption{Special cases of the GS model with corresponding parameters}\label{specialcase_table}
\begin{tabular}{cc}
\toprule
Special case& Parameter value\\
\midrule
bursty IN& $\rho=0$ and $\tilde{\Sigma}\neq\mathbf{I}_p$\\
WGN& $p=1$ or $\rho=1$\\
white IN& $\rho=0$ and $p=1$\\
white mixed noise& $p=1$\\
\bottomrule
\end{tabular}
\end{center}
\end{table}
\end{property}

The Property \ref{property_4} can be verified by plugging the parameters in \eqref{bursty_mixed_noise_model}. Note that `white' in Table \ref{specialcase_table} denotes that the noise samples are mutually independent. Besides, we can also consider the bursty mixed noise as WGN and pure bursty IN when $\gamma_g\gg\gamma_s$ and $\gamma_s\gg\gamma_g$, respectively. Then, the marginal distribution of \eqref{bursty_mixed_noise_model} is investigated, which can be applied to derive the moment property of the GS model.
\begin{property}\label{property_5}
Let the random vector $\mathbf{N}=[N_1,N_2,\cdots,N_p]^{\top}$ follows the GS model. Then, the component $N_j(j=1,2,\cdots,p)$ are identically distributed and the PDF of $N_1$ in \eqref{bursty_mixed_noise_model} is as follows,
\begin{align}\label{marginal_distribution_GS}
f_M(n)=&\frac{\rho}{2\sqrt{\pi}\gamma_g}\exp\left(-\frac{n^2}{4\gamma_g^2}\right)+\frac{(1-\rho)k_m}{(\alpha+(n\Sigma^{-1/2}_p)^2)^{\frac{\alpha+1}{2}}},
\end{align}
where
\begin{align}
k_m=\frac{\alpha^{\frac{\alpha}{2}}\Sigma^{-1/2}_p\Gamma(\frac{\alpha+1}{2})}{2\Gamma(\frac{3}{2})\Gamma(\frac{\alpha}{2})}
\end{align}
and $\Sigma^{-1/2}_p\triangleq\Sigma^{-1/2}(p,p)$.
\end{property}
\begin{IEEEproof}
The proof is relegated to Appendix \ref{appendix_property_5}.
\end{IEEEproof}

Based on the Property \ref{property_5}, it can be seen that the component of the GS random vector is also a heavy-tailed distribution with the same characteristic parameter $\alpha$ and the weight parameter $\rho$. The univariate distribution is the same with the special case of \eqref{bursty_mixed_noise_model} with $p=1$. Meanwhile, the marginal PDF is not correlated to the covariance matrix in the original multivariate distribution. Then, we can also obtain the moment property with the marginal PDF and it is difficult to derive from the CF due to the lack of the analytic expression.
\begin{property}\label{property_6}
Let the random vector $\mathbf{N}=[N_1,N_2,\cdots,N_p]^{\top}$ follows the GS model. Then, the $p$-th moment of $N_1$ is finite if $r<\alpha$ and
\begin{align}\label{moment}
\mathbb{E}_{\mu_{N_1}}[|N_1|^r]=&\frac{(2\gamma_g)^{r}\rho}{\sqrt{\pi}}\Gamma\left(\frac{r+1}{2}\right)\nonumber\\
&+\frac{(1-\rho)\alpha^{\frac{r}{2}}\Gamma(\frac{r+3}{2})\Gamma(\frac{a-r}{2})\left(\Sigma^{-1/2}_p\right)^{r}}{\Gamma(\frac{3}{2})\Gamma(\frac{\alpha}{2})(r+1)}.
\end{align}
\end{property}
\begin{IEEEproof}
The proof is relegated to Appendix \ref{appendix_property_6}.
\end{IEEEproof}

Based on the property \ref{property_6}, we have $\mathbb{E}_{\mu_{N_1}}[|N_1|]=\frac{2\gamma_g\rho}{\sqrt{\pi}}+\frac{(1-\rho)\sqrt{\alpha}\Gamma(\frac{a-1}{2})\Sigma^{-1/2}_p}{2\Gamma(\frac{3}{2})\Gamma(\frac{\alpha}{2})}$ and therefore, the $\hat{\alpha}$ can be estimated if the $\hat{\rho}$, $\hat{\gamma}_g$ and $\hat{\Sigma}$ have been determined.

\section{Parameter estimation of the proposed GS model}\label{parameterestimation}
In this section, we design the parameter estimation algorithm of our proposed GS model. Since there are several parameters Besides, the estimation performance and convergence will be discussed.

\subsection{Estimation of $\tilde{\Sigma}$ and $p$}
The correlation between noise samples is presented by the regularized covariance matrix $\tilde{\Sigma}$ and its dimension $p$, which reflects the length of impulsive cluster. Without loss of generality, we suppose that the noise sample $n(i)$ and $n(i+\Delta)$ are mutually independent for $\Delta>p$. We adopt the sample covariance matrix $\Sigma_S$ as the estimator of the $\tilde{\Sigma}$. It is given as follows \cite{paper19},
\begin{equation}\label{estimation_sigma}
\Sigma_S=\frac{1}{L-1}\sum_{i=1}^{L}(\mathbf{n}_i-\boldsymbol{\mu})(\mathbf{n}_i-\boldsymbol{\mu})^\top,
\end{equation}
where $\mathbf{n}_i$ is the $i$-th sample vector, $L$ is the number of vector samples, and the mean vector  $\boldsymbol{\mu}=\sum_{i=1}^{L}\mathbf{n}_i/L$. $\mathbf{n}_i$ is generated from the noise sequence with a sliding window of size $p$. Let $L_N$ be the length of noise sample sequence and the noise sequence
\begin{equation}
[\underbrace{n(1),\cdots,n(p)}_{\mathbf{n}_1},\overbrace{n(p+1),\cdots,n(2p)}^{\mathbf{n}_{p+1}},\cdots,n(L_N)]^\top \nonumber
\end{equation}
contains $L_N-p+1$ noise vector samples.

To determine $p$, we calculate the samples correlation $Cov(N(i),N(i+\Delta))$ as follows,
\begin{equation}
\begin{aligned}
&Cov(N(i),N(i+\Delta))\\
&=\frac{1}{L_N-\Delta-1}\sum_{i=1}^{L_N-\Delta}(n(i)-\mu)(n(i+\Delta)-\mu)
\end{aligned}
\end{equation}
where $n(i)$ is the $i$-th noise sample. The sample mean $\mu=\sum_{i=1}^{L_N}n(i)/L_N$.
We set a threshold $T$ and consider $n(i)$ and $n(i+\Delta)$ to be independent if $|Cov(N(i),N(i+\Delta))|<T$. Then, $p$ is estimated as follows,
\begin{equation}\label{estimation_p}
\hat{p}=\min\left\{k-1:\frac{|Cov(N(i),N(i+k))|}{|Cov(N(i),N(i))|}<T,k\in\mathbb{Z}^+\right\}.
\end{equation}

Note that the mixed noise samples are considered to be approximately white for $\frac{|Cov(N(i),N(i+1))|}{|Cov(N(i),N(i))|}<T$. To avoid confusion, we use $\hat{\Sigma}_R$ instead of $\hat{\tilde{\Sigma}}$ to represent the estimation of $\tilde{\Sigma}$. Notice that the diagonal element of $\tilde{\Sigma}$ is 1 and therefore,
\begin{equation}\label{estimation_sigma_regular}
\hat{\Sigma}_R=\frac{\Sigma_S}{\Sigma_S(1,1)}.
\end{equation}

\subsection{Estimation of $\rho$ and $\gamma_g$}
The weight factor $\rho$ of GS model denotes the ratio of WGN and IN. The bursty mixed noise following GS model will degrade to WGN when $\rho=1$ or $\gamma_g\rightarrow+\infty$. ${\rho}$ and ${\gamma_g}$ can be estimated based on the CF of GS model. Before proceeding to describe the estimation method, we first give the following theorem.

\begin{theorem}\label{theorem_1}
Let $C$ be a $p\times p$ real symmetric matrix and $C\neq \mathbf{I}_p$. There exist $\mathbf{t}_i\in\mathbb{R}^{p\times 1}$ and $\mathbf{t}_j\in\mathbb{R}^{p\times 1}$ such that $\Vert\mathbf{t}_i\Vert\neq\Vert\mathbf{t}_j\Vert$ and $\Vert C^{1/2}\mathbf{t}_i\Vert=\Vert C^{1/2}\mathbf{t}_j\Vert$.
\end{theorem}
\begin{IEEEproof}
The proof is relegated to Appendix \ref{appendix_theorem_1}.
\end{IEEEproof}

It follows theorem \ref{theorem_1} and \eqref{CF_GS} that there exist $\mathbf{t}_i$ and $\mathbf{t}_j$ such that
\begin{equation}
\Phi_M(\mathbf{t}_i)-\Phi_M(\mathbf{t}_j)=\rho \left(\exp\left(-\gamma_g^2\Vert\mathbf{t}_i\Vert^2\right)-\exp\left(-\gamma_g^2\Vert\mathbf{t}_j\Vert^2\right)\right).
\end{equation}

Then, we can obtain the estimation $\hat{\rho}$ of ${\rho}$ and $\hat{\gamma}_g$ of ${\gamma}_g$ based on the following equations:
\begin{equation}\label{estimation_gama_g}
\frac{\Phi_M(\mathbf{t}_3)-\Phi_M(\mathbf{t}_2)}{\Phi_M(\mathbf{t}_2)-\Phi_M(\mathbf{t}_1)}=\underbrace{\frac{\exp(-\hat{\gamma}_g^2(\Vert\mathbf{t}_3\Vert^2-\Vert\mathbf{t}_2\Vert^2))-1}{1-\exp(-\hat{\gamma}_g^2(\Vert\mathbf{t}_1\Vert^2-\Vert\mathbf{t}_2\Vert^2))}}_{\triangleq g(\hat{\gamma}_g)},
\end{equation}
\begin{equation}\label{estimation_rho}
\hat{\rho}=\frac{\Phi_M(\mathbf{t}_3)-\Phi_M(\mathbf{t}_2)}{\exp\left(-\hat{\gamma}_g^2\Vert\mathbf{t}_3\Vert^2\right)-\exp\left(-\hat{\gamma}_g^2\Vert\mathbf{t}_2\Vert^2\right)}. \end{equation}

As \eqref{estimation_gama_g} does not admit analytic solution, it has to be solved via numerical methods. The choice of $\mathbf{t}_i,i=1,2,3$ is also provided in Appendix \ref{appendix_theorem_1}. Meanwhile, $\Phi_M(\mathbf{t})$ can be approximated by the empirical characteristic function (ECF) $\tilde{\Phi}_M(\mathbf{t})$, i.e.,
\begin{equation}\label{emperical_CF}
\tilde{\Phi}_M(\mathbf{t})=\frac{1}{L}\sum_{i=1}^{L}\exp(j\mathbf{t}^\top\mathbf{n}_i).
\end{equation}
where $\mathbf{n}_i$ represents the $i$-th noise vector sample. For the ECF \eqref{emperical_CF}, we have the following convergence theorem.
\begin{theorem}\label{theorem_2}
Assume that $\mathbf{X}$ is a multivariate random vector with covariance matrix $\Sigma\in\mathbb{S}^{M}_{++}$ and $M<+\infty$. Let the CF of $\mathbf{X}$ to be $\Phi_M(\mathbf{t})$ and $\tilde{\Phi}_M(\mathbf{t})$ is defined in \eqref{emperical_CF}. Then, $\tilde{\Phi}_M(\mathbf{t})\stackrel{L^2}{\longrightarrow}\Phi_M(\mathbf{t})$ where the notation `$\stackrel{L^2}{\longrightarrow}$' represents the convergence in $L^2$.
\end{theorem}
\begin{IEEEproof}
The proof is relegated to Appendix \ref{appendix_theorem_2}.
\end{IEEEproof}

Based on the theorem \ref{theorem_2}, we can also immediately obtain that $\tilde{\Phi}_M(\mathbf{t})\stackrel{p}{\longrightarrow}\Phi_M(\mathbf{t})$ where `$\stackrel{p}{\longrightarrow}$' denotes convergence in probability since $\forall\epsilon>0$,
\begin{align}
\lim\limits_{n\rightarrow+\infty}P&\left(|\tilde{\Phi}_M(\mathbf{t})-\Phi_M(\mathbf{t})|>\epsilon\right)\nonumber\\
&\leq\lim\limits_{n\rightarrow+\infty}\frac{\mathbb{E}_{\mu_{\mathbf{X}}}(|\tilde{\Phi}_M(\mathbf{t})-\Phi_M(\mathbf{t})|^2)}{\epsilon^2}=0.
\end{align}

It follows theorem \ref{theorem_2} that $\mathbb{E}_{\mu_{\mathbf{X}}}(\tilde{\Phi}_M(\mathbf{t}))=\Phi_M(\mathbf{t})$. Based on \eqref{estimation_rho}, we have
\begin{equation}\label{rho_equation}
\hat{\rho}\left[\exp\left(-\hat{\gamma}_g^2\Vert\mathbf{t}_3\Vert^2\right)-\exp\left(-\hat{\gamma}_g^2\Vert\mathbf{t}_2\Vert^2\right)\right]
=\tilde{\Phi}_M(\mathbf{t}_3)-\tilde{\Phi}_M(\mathbf{t}_2).
\end{equation}

By applying the expectation of \eqref{rho_equation}, we have
\begin{align}
\mathbb{E}_{\mu_{\mathbf{X}}}\big(\hat{\rho}\big[\exp\left(-\hat{\gamma}_g^2\Vert\mathbf{t}_3\Vert^2\right)-&\exp\left(-\hat{\gamma}_g^2\Vert\mathbf{t}_2\Vert^2\right)\big]\big)\nonumber\\
&=\mathbb{E}_{\mu_{\mathbf{X}}}\left(\tilde{\Phi}_M(\mathbf{t}_3)-\tilde{\Phi}_M(\mathbf{t}_2)\right)\nonumber\\
&=\Phi_M(\mathbf{t}_3)-\Phi_M(\mathbf{t}_2).
\end{align}

Consequently,
\begin{equation}
\mathbb{E}_{\mu_{\mathbf{X}}}(\hat{\rho})=\frac{\Phi_M(\mathbf{t}_3)-\Phi_M(\mathbf{t}_2)}{\exp\left(-\hat{\gamma}_g^2\Vert\mathbf{t}_3\Vert^2\right)-\exp\left(-\hat{\gamma}_g^2\Vert\mathbf{t}_2\Vert^2\right)}=\rho
\end{equation}

Similarly, it can be verified that $\hat{\gamma}_g$ is an unbiased estimation of ${\gamma}_g$. Subsequently, we give the following lemma to show the monotonicity of $g(\gamma_g)$. It can be leveraged to reduce the complexity of numerical method.
\begin{lemma}\label{lemma_2}
Define
\begin{equation}
g(x)=\frac{\exp\left(ax^2\right)-1}{1-\exp\left(bx^2\right)},x>0.
\end{equation}

Then, $g(x)$ is a monotone function for $\forall a\in \mathbb{R},b\in \mathbb{R}-\{0\}$.
\end{lemma}
\begin{IEEEproof}
The derivative of $g(x)$ is as follows,
\begin{align}
\frac{\partial g(x)}{\partial x}=\frac{2\exp((b-a)x^2)x}{(\exp(bx^2)-1)^2}&[a(1-\exp(bx^2))\nonumber\\
&+b(\exp(ax^2)-1)]
\end{align}

Thus, the sign of the derivative is determined by $[a(1-\exp(bx^2))+b(\exp(ax^2)-1)]$. For $a>0$ and $b>0$, we express this part by the Maclaurin's series as
\begin{align}\label{derivative_of_g}
&a(1-\exp(bx^2))+b(\exp(ax^2)-1)\nonumber\\
=&-a\sum_{k=1}^{+\infty}\frac{(bx^2)^k}{k!}+b\sum_{k=1}^{+\infty}\frac{(ax^2)^k}{k!} \nonumber\\
=&ab\sum_{k=2}^{+\infty}\frac{x^{2k}}{k!}\left(a^{k-1}-b^{k-1}\right)
\end{align}

It is easy to observe that $g(x)$ is a monotone function from \eqref{derivative_of_g} since $\frac{\partial g(x)}{\partial x}$ does not change sign on the feasible domain. The analysis of the case for $a<0$ and $b<0$ is equivalent to \eqref{derivative_of_g}. It is obvious that $g(x)$ is monotonic when we have $ab<0$ and $g(x)$ degrades to constant if $a=b$ or $a=0$.
\end{IEEEproof}

It follows lemma \ref{lemma_2} that $g(\hat{\gamma}_g)$ is a monotone function and therefore, some low-complexity algorithms, e.g., the binary search, can be used to obtain $\hat{\gamma}_g$.

\subsection{Estimation of $\alpha$ and $\gamma_s$}
It is challenging to estimate $\alpha$ and $\gamma_s$ since the CF is quite complex and there are no statistics with simple expressions, e.g., moments, quantile, etc. Moreover, directly joint estimation of $\alpha$ and $\gamma_s$ via ML method is not appropriate due to prohibitively high computation complexity. Meanwhile, some conventional DOF estimators for multivariate student distribution also could not be utilized. For instance, the Hill estimator is a biased estimator based on the order statistics of the student distribution \cite{paper16}. As the derivation is related to the asymptotic behaviors of student distribution, the number of order statistic used to estimate DOF is difficult to choose. In fact, there is a tradeoff between the mean bias and estimation variance. Therefore, we herein combine the PDF at the origin and ML estimator to obtain the estimation $\hat{\alpha}$ of ${\alpha}$ and $\hat{\gamma}_s$ of ${\gamma}_s$. Specifically, based on \eqref{bursty_mixed_noise_model}, we have
\begin{equation}
f_M(\mathbf{0})=\frac{\rho}{(\sqrt{2}\pi\gamma_g)^p}+\frac{(1-\rho)\Gamma((\alpha+p)/2)}{\Gamma(\alpha/2)(2\gamma_s^2\alpha\pi)^{p/2}\sqrt{\det(\tilde{\Sigma})}}.
\end{equation}

Therefore, based on the estimation of $\hat{\Sigma}_R$, $\hat{p}$, $\hat{\rho}$ and $\hat{\gamma}_g$, we can get that
\begin{align}\label{estimation_gamma_s}
\hat{\gamma}_s=\Bigg\{&\frac{(1-\hat{\rho})\Gamma((\alpha+\hat{p})/2)}{\Gamma(\alpha/2)(2\alpha\pi)^{\hat{p}/2}\sqrt{\det(\hat{\Sigma}_R)}}\nonumber\\
&\times\left[f_M(\mathbf{0})-\frac{\hat{\rho}}{(\sqrt{2}\pi\hat{\gamma}_g)^{\hat{p}}}\right]^{-1}\Bigg\}^{1/\hat{p}}.
\end{align}
where $f_M(\mathbf{0})$ can be calculated based on the noise sample vectors by the Gaussian kernel estimation (GKE), i.e.,
\begin{equation}\label{GKE}
f_M(\mathbf{0})=\frac{1}{L}\sum_{i=1}^{L}\frac{1}{(\sqrt{2\pi}h)^{\hat{p}}}\exp\left(-\frac{\mathbf{n}_i^2}{2h^2}\right),
\end{equation}
where $h$ is the hyper-parameter to be determined empirically. With the aid of \eqref{estimation_gamma_s}, the ML estimation can be applied since the unknown $\gamma_s$ can be considered as a function of $\alpha$ and the other parameters which have been determined. The ML estimator can be expressed as
\begin{align}\label{estimation_alpha}
\hat{\alpha}=\max\limits_{\alpha}\sum_{i=1}^{L}\log [f_M(\mathbf{n}_i)].
\end{align}

However, it is difficult to find the closed-form expression of $\hat{\alpha}$. Note that $\alpha$ is the single unknown parameter after plugging \eqref{estimation_gamma_s} into \eqref{estimation_alpha}. Some low-complexity numerical algorithms, such as one-dimensional grid search or gradient descent, can be used.

Finally, the proposed parameter estimation procedure is summarized as Algorithm. \ref{algorithm_procedure}. Note that $\mathbf{diag}(\cdot)$ represents a diagonal matrix.
\begin{algorithm}
\caption{Parameter estimation algorithm for the GS model}\label{algorithm_procedure}
\KwData{Bursty mixed noise sample vectors $\mathbf{n}_i,i=1,2,$ $\cdots,L$, covariance threshold $T$}
\KwResult{Estimated parameters $\hat{\Sigma}_R$, $\hat{p}$, $\hat{\rho}$, $\hat{\gamma}_g$, $\hat{\alpha}$, $\hat{\gamma}_s$}
$k=1$\;
\While{$\frac{|Cov(N(i),N(i+k))|}{|Cov(N(i),N(i))|}\geq T$}
{$k\leftarrow k+1$\;}
$\hat{p}\leftarrow k-1$\;
Estimate $\tilde{\Sigma}$ based on \eqref{estimation_sigma} and \eqref{estimation_sigma_regular}\;
Perform eigenvalue decomposition to get $\hat{\Sigma}_R=P^\top \mathbf{diag}($$\lambda_1,\cdots,\lambda_p)P$\;
$\mathbf{t}_1=P[1,\sqrt{\lambda_3/\lambda_2},\sqrt{\lambda_2/\lambda_3},1,\cdots,1]^\top,\mathbf{t}_2=P[1,1,$ $1,\cdots,1]^\top,\mathbf{t}_3=P[\sqrt{\lambda_2/\lambda_1},\sqrt{\lambda_1/\lambda_2},1,\cdots,1]^\top$\;
Calculate $\tilde{\Phi}_M(\mathbf{t}_k)=\frac{1}{L}\sum_{i=1}^{L}\exp(j\mathbf{t}_k^\top\mathbf{n}_i),k=1,2,3$\;
Solve $\eqref{estimation_gama_g}$ by binary search to obtain $\hat{\gamma}_g$\;
Plug $\hat{\gamma}_g$ in \eqref{estimation_rho} to obtain $\hat{\rho}$\;
Solve $\eqref{estimation_alpha}$ by grid search to obtain $\hat{\alpha}$\;
Plug $\hat{\Sigma}_R$, $\hat{p}$, $\hat{\rho}$, $\hat{\gamma}_g$, $\hat{\alpha}$ in \eqref{estimation_gamma_s} to obtain $\hat{\gamma}_s$.
\end{algorithm}

\section{simulation}\label{simulation}
In this section, experimental results are carried out under various noise scenarios to validate the ability of the proposed model to describe the bursty mixed noise and the performance of parameter estimation. According to property \ref{property_6}, the variance of the GS model does not exist and thus, the conventional signal-to-noise ratio (SNR) could not be used in simulation. Here, we define the generalized signal-to-noise ratio (GSNR) as follows,
\begin{equation}\label{GSNR}
\text{GSNR(dB)}=10\log_{10}\frac{P_s}{2(\gamma_g^2+\gamma_s^{\alpha})},
\end{equation}
where the $P_s$ is the power of the transmitted signals.

\subsection{Model verification}
In this subsection, we generate the noise samples following the proposed model and compare it with other models to verify the performance and generality of the GS model. The bursty mixed noise with memory following \eqref{bursty_mixed_noise_model} can be generated by Acceptance-Rejection sampling method. In simulation, we set the horizontal axis range of PDF to $[-500,500]$ since there are numerous outliers, especially when $\alpha$ and $\rho$ is smaller. To analyze the influence of $\alpha$, $\rho$ and $p$ on the feature of the mixed noise in the time domain. Therefore, we fix that $\gamma_s=\gamma_g=2$. The $\alpha$ are chosen to be 1.2 and 1.8 to separately represent the noise with larger impulsity and the noise more close to WGN. The regularized covariance matrix is set to be
\begin{equation}\label{covariance_simulation_2}
p=2:\tilde{\Sigma}=
\left[
\begin{array}{cc}
  1 & 0.7 \\
  0.7 & 1
\end{array}
\right],
\end{equation}
\begin{equation}\label{covariance_simulation_5}
p=5:\tilde{\Sigma}=
\left[
\begin{array}{ccccc}
  1 & 0.8 &0.6 &0.4 &0.2 \\
  0.8 & 1 &0.8 &0.6 &0.4 \\
  0.6 & 0.8 &1 &0.8 &0.6 \\
  0.4 & 0.6 &0.8 &1 &0.8 \\
  0.2 & 0.4 &0.6 &0.8 &1
\end{array}
\right].
\end{equation}
\begin{figure*}[htbp]
\centering
\subfloat[$\alpha=1.2$, $\rho=0$, $p=5$]{\includegraphics[width=1.7in]{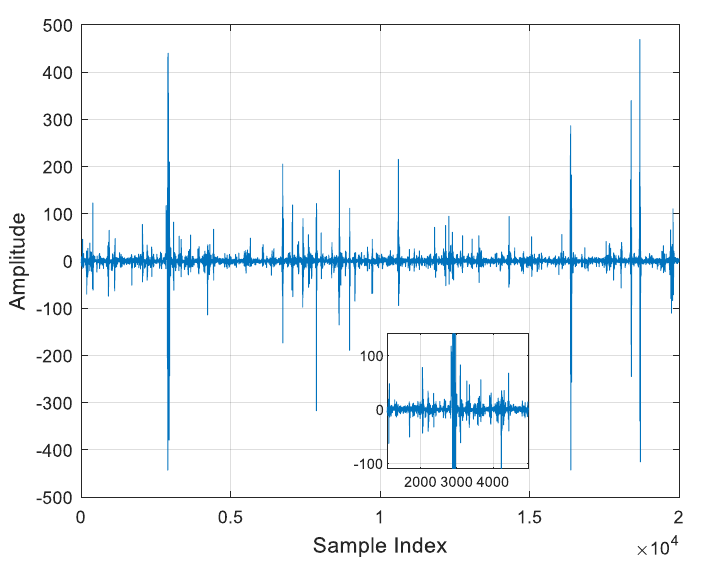}\label{fig_1_a}}
\subfloat[$\alpha=1.2$, $\rho=0.5$, $p=5$]{\includegraphics[width=1.7in]{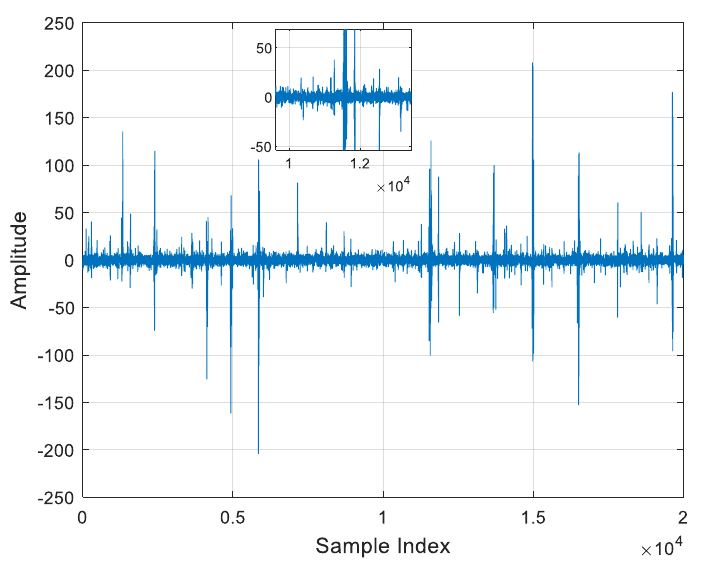}\label{fig_1_b}}
\subfloat[$\alpha=1.2$, $\rho=0.8$, $p=5$]{\includegraphics[width=1.7in]{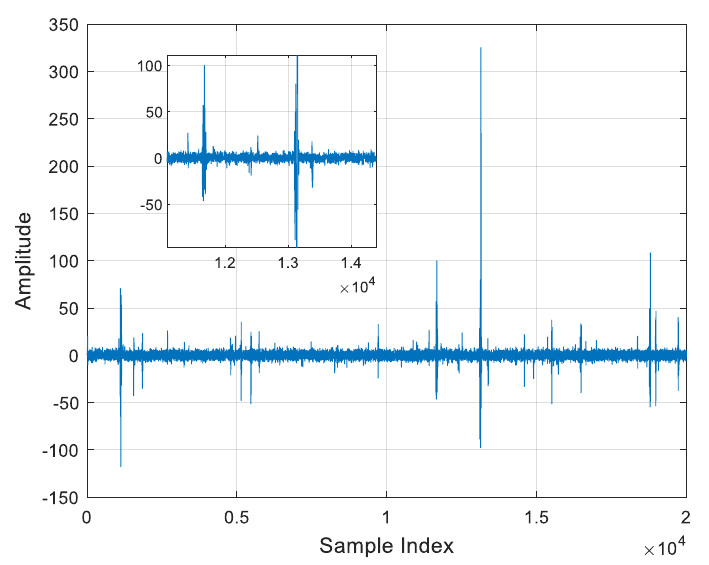}\label{fig_1_c}}
\subfloat[$\alpha=1.2$, $\rho=1$, $p=5$]{\includegraphics[width=1.7in]{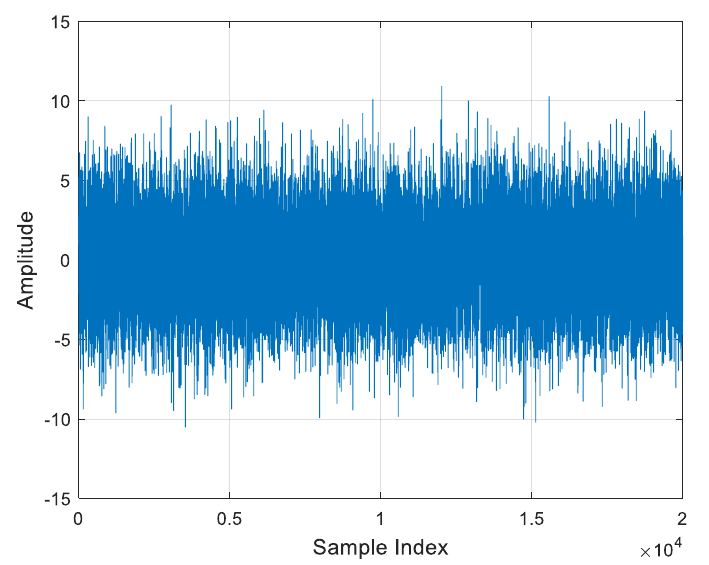}\label{fig_1_d}}

\subfloat[$\alpha=1.8$, $\rho=0$, $p=5$]{\includegraphics[width=1.7in]{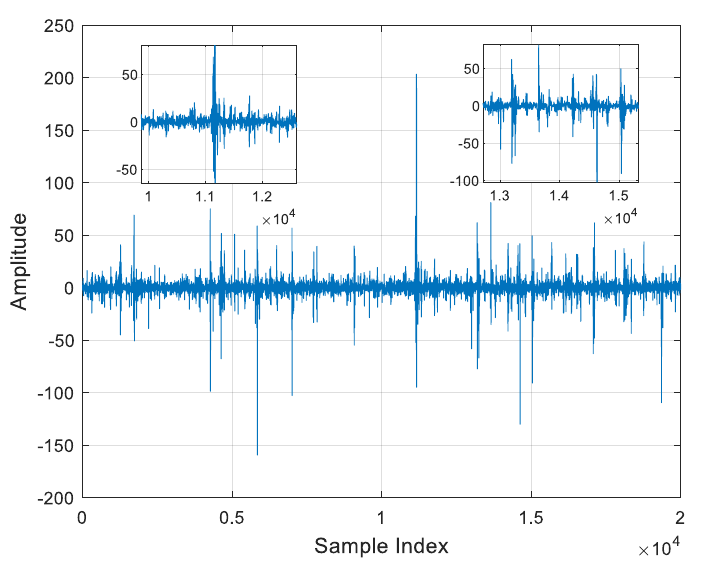}\label{fig_1_e}}
\subfloat[$\alpha=1.8$, $\rho=0.5$, $p=5$]{\includegraphics[width=1.7in]{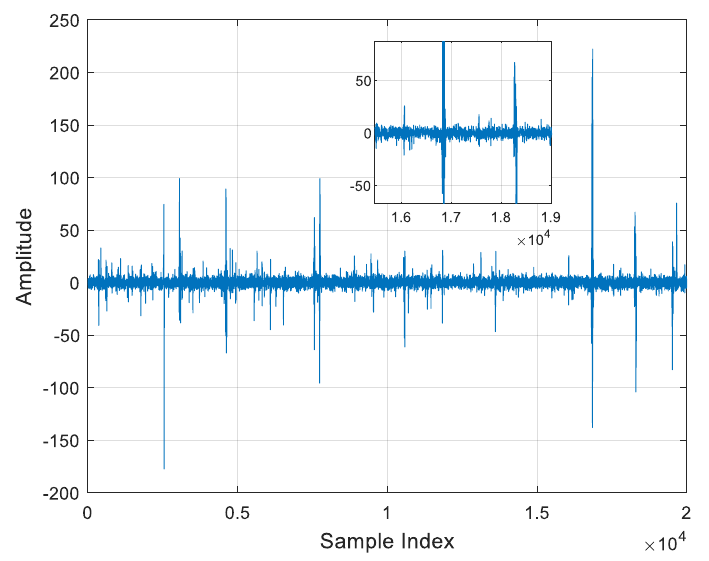}\label{fig_1_f}}
\subfloat[$\alpha=1.8$, $\rho=0$, $p=2$]{\includegraphics[width=1.7in]{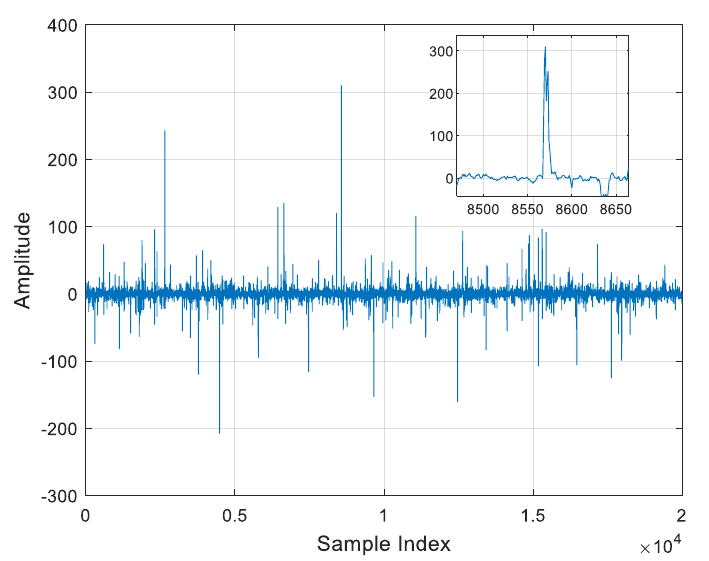}\label{fig_1_g}}
\subfloat[$\alpha=1.8$, $\rho=0.5$, $p=2$]{\includegraphics[width=1.7in]{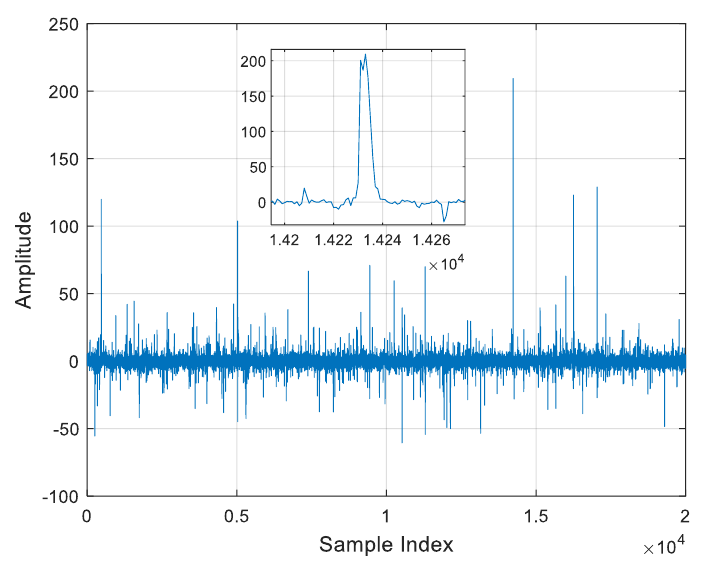}\label{fig_1_h}}
\caption{Bursty mixed noise generated by the GS model}
\label{fig_1}
\end{figure*}

The bursty mixed noise generated by the GS model is shown in Fig. \ref{fig_1}. It can be seen that the generated noise is composed of background WGN and the bursty IN with different cluster lengths. From Fig. \ref{fig_1_a}-Fig. \ref{fig_1_c}, it is obvious that $\rho$ is tightly related to the impulsity of the noise, which is reflected the number of IN cluster and the power of background WGN. Note that the noise becomes WGN when $\rho=1$ and is closed to the noise following $\alpha$SG model when $\rho=0$, which is demonstrated by Fig. \ref{fig_1_d} and \ref{fig_1_a}, respectively. Fig. \ref{fig_1_b} and Fig. \ref{fig_1_f} show that larger $\alpha$ corresponds to smaller impulsity since there are fewer outliers and their amplitude is relatively smaller. Fig. \ref{fig_1_f}-Fig. \ref{fig_1_h} verify that the memory order $p$ is linked to the IN cluster length and the bursty property will diminish if $p=1$.

\begin{figure}[htbp]
\centering
\includegraphics[width=3.2in]{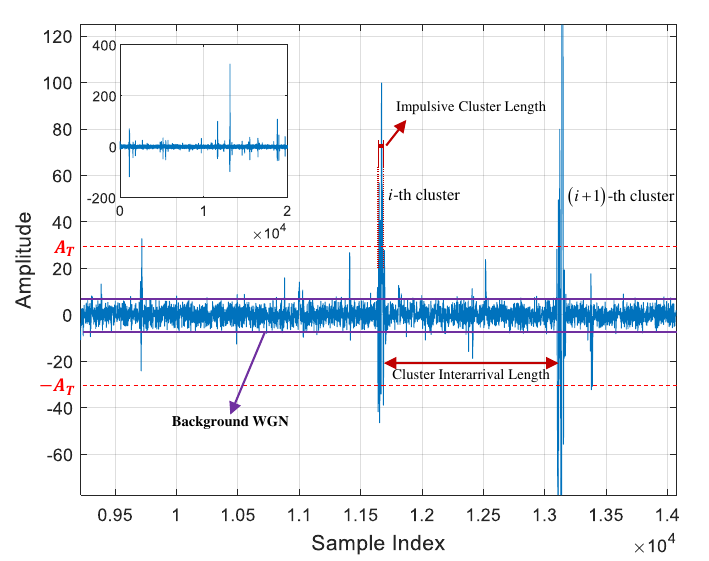}
\caption{The feature parameters of bursty mixed noise}
\label{fig_feature_parameter}
\end{figure}

Then, we describe features of the bursty mixed noise by Fig. \ref{fig_feature_parameter} before the comparison between different noise models is carried out. The similarity between two PDFs is usually measured by their KL divergence. However, it is difficult to calculate KL divergence due to the high dimension of noise PDF and therefore, we compare the similarity of the noise from 3 aspects: amplitude, impulsive cluster length (ICL) and the cluster interarrival length (CIL). Fig. \ref{fig_feature_parameter} further describes these feature parameters clearly. We define that the noise sample with amplitude larger than the threshold $A_T$ is a impulsive cluster. The ICL is the number of consecutive noise samples whose amplitudes are greater than $A_T$. CIL is the number of noise samples between the ending of the $i$-th and the starting of the $(i+1)$-th impulsive cluster.

Then, we compare the noise based on different popular models and the real sferic noise data measured in \cite{paper4}. The MM model is a modified version of the MG model and therefore, we mainly consider the $\alpha$SG and MM model as the baseline. The measured noise data could not be obtained and herein is generated by the Poisson cluster (PC) model built in \cite{paper4} since the model can well match the real noise data. To avoid redundancy, we herein only consider different $\alpha$ and set $\gamma_s=\gamma_g=2$, $\rho=0.6$ and $p=5$. The memory order and covariance matrix in \eqref{covariance_simulation_5} is applied.

\begin{figure*}[htbp]
\centering
\subfloat[$\alpha=1.2$]{\includegraphics[width=3in]{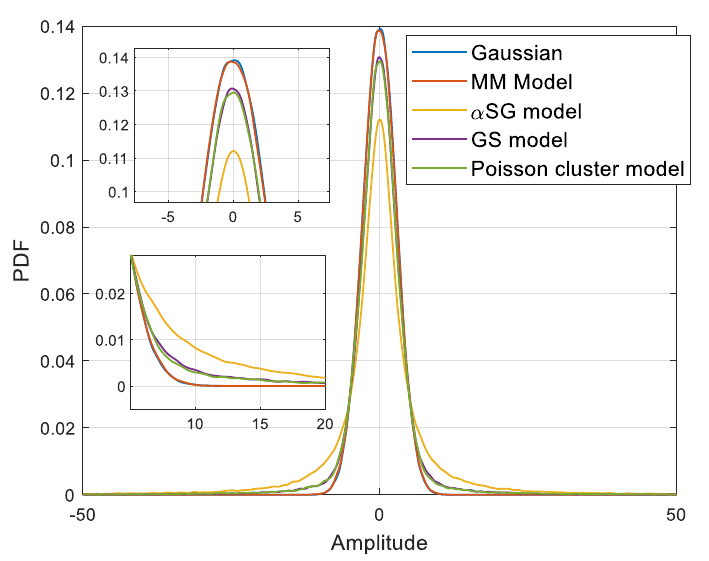}\label{fig_2_a_1}}
\subfloat[$\alpha=1.8$]{\includegraphics[width=3in]{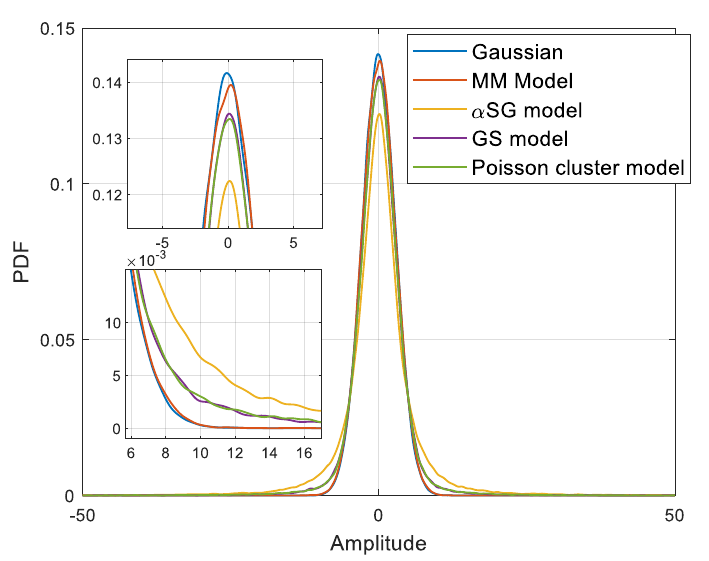}\label{fig_2_a_2}}
\caption{The amplitude PDF comparison of noise based on various models}
\label{fig_2_a}
\end{figure*}

\begin{figure*}[htbp]
\centering
\subfloat[$\alpha=1.2$]{\includegraphics[width=3in]{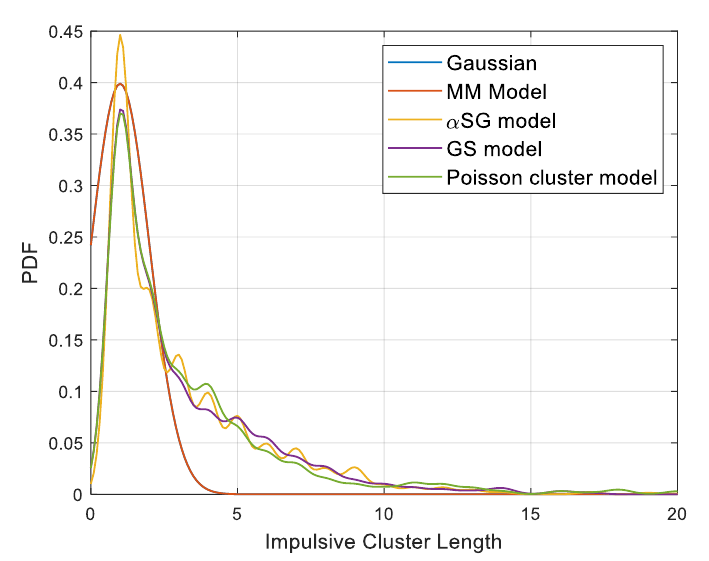}\label{fig_2_b_1}}
\subfloat[$\alpha=1.8$]{\includegraphics[width=3in]{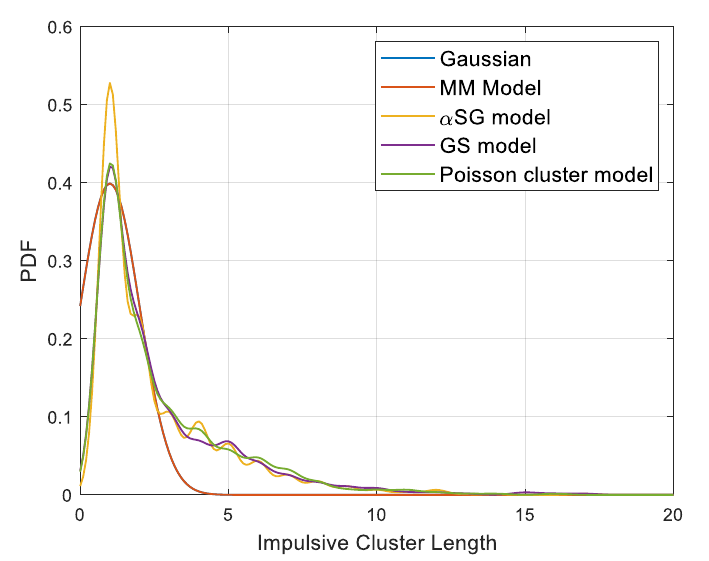}\label{fig_2_b_2}}
\caption{The impulsive cluster length PDF comparison of noise based on various models. The PDFs of Gaussian could not be observed because it equals to zero.}
\label{fig_2_b}
\end{figure*}

\begin{figure*}[htbp]
\centering
\subfloat[$\alpha=1.2$]{\includegraphics[width=3in]{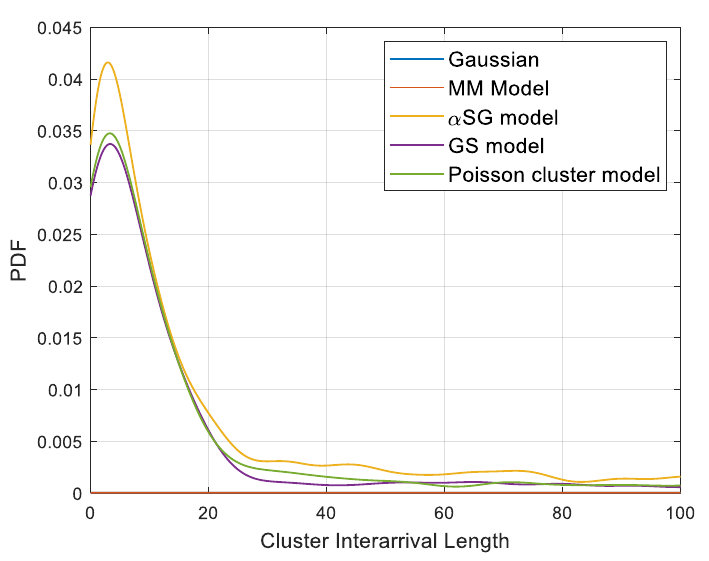}\label{fig_2_c_1}}
\subfloat[$\alpha=1.8$]{\includegraphics[width=3in]{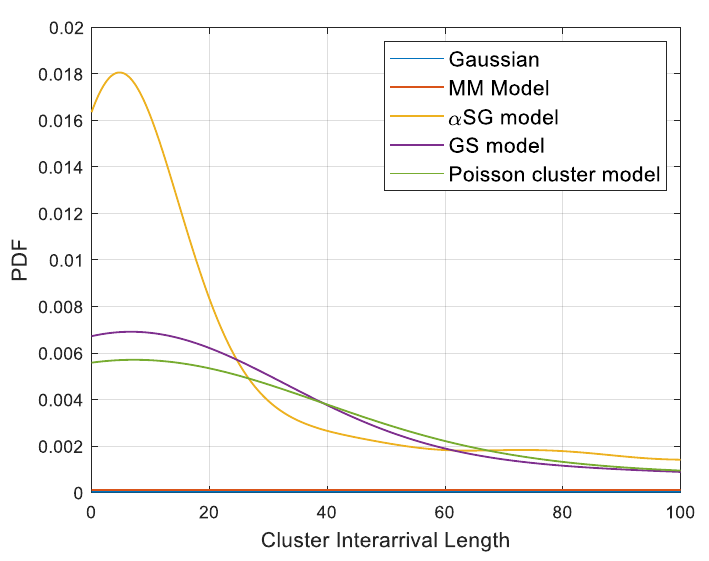}\label{fig_2_c_2}}
\caption{The cluster interarrival length PDF comparison of noise based on various models. The PDFs of Gaussian and MM model are vague because they are quite close to the horizontal axis.}
\label{fig_2_c}
\end{figure*}

Comparison results are presented in Fig \ref{fig_2_a}-\ref{fig_2_c}. According to the results, the curves of the GS model fit the PC model tightly, which demonstrates our model describes the bursty mixed noise very well. In Fig. \ref{fig_2_a}, the tail section of $\alpha$SG distribution is heavier than the GS model since $\alpha$SG model does not take into account the Gaussian component. For the same reason, the tail section of the ICL PDF of $\alpha$SG model is similar to our model but with more oscillation and concentrates more on the region close to the zero. Besides, the CIL of noise based on $\alpha$SG distribution is smaller than the GS and PC model since the power of WGN is converted to a more impulsive cluster, which mismatches the reality. The curve of Gaussian distribution in Fig. \ref{fig_2_b} and \ref{fig_2_c} could not be observed because the amplitude of Gaussian noise is smaller than the threshold and therefore, there is no ICL for Gaussian model or its ICL can be considered as infinite. Fig. \ref{fig_2_b} and \ref{fig_2_c} also verify that the MM model does not admit the heavy tail characteristic.

\subsection{Parameter estimation performance}

\begin{figure*}[htbp]
\centering
\subfloat[Estimation of $\tilde{\Sigma}$]{\includegraphics[width=2.4in]{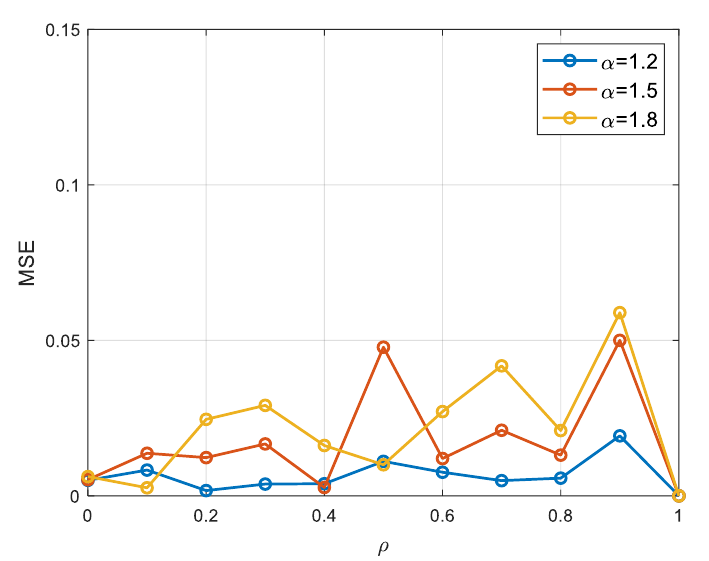}\label{fig_3_a}}
\subfloat[Estimation of $\rho$]{\includegraphics[width=2.4in]{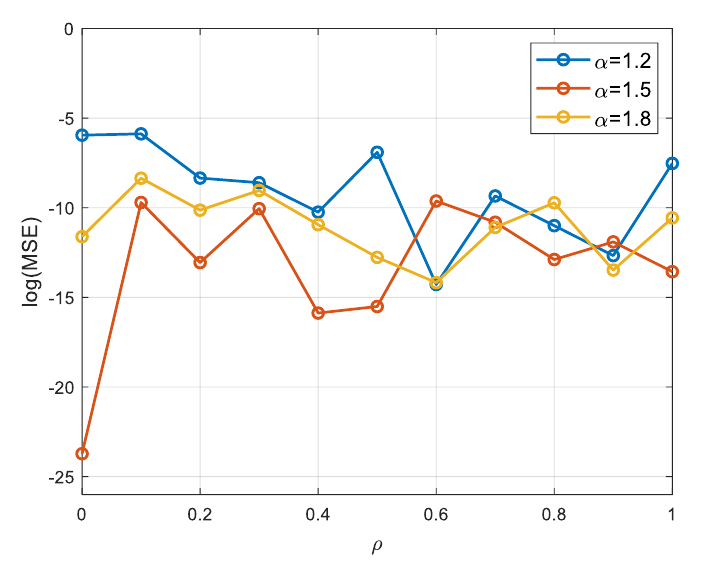}\label{fig_3_b}}

\subfloat[Estimation of $\gamma_g$]{\includegraphics[width=2.4in]{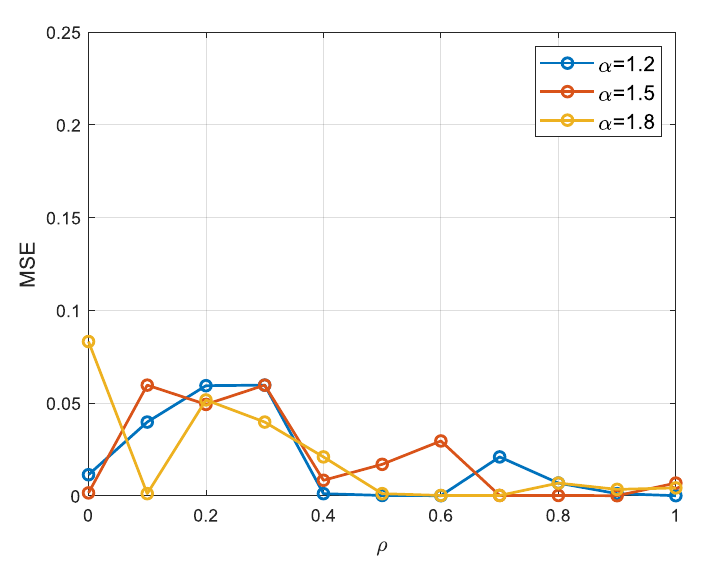}\label{fig_3_c}}
\subfloat[Estimation of $\gamma_s$]{\includegraphics[width=2.4in]{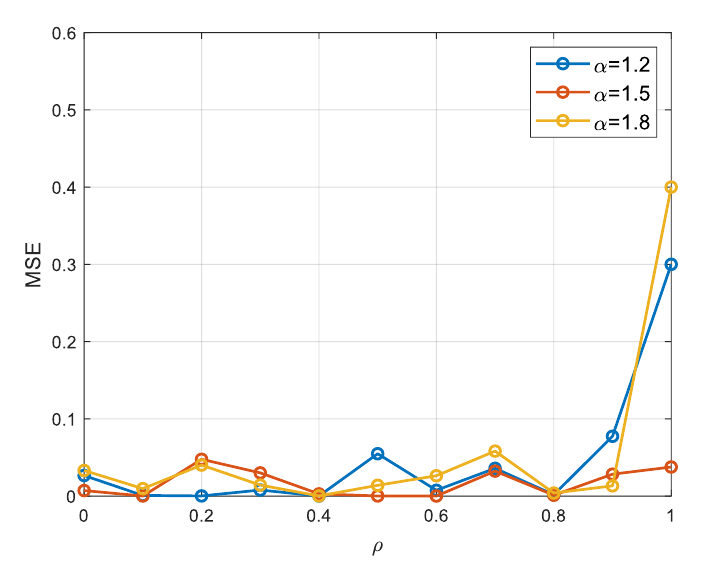}\label{fig_3_d}}
\subfloat[Estimation of $\alpha$]{\includegraphics[width=2.4in]{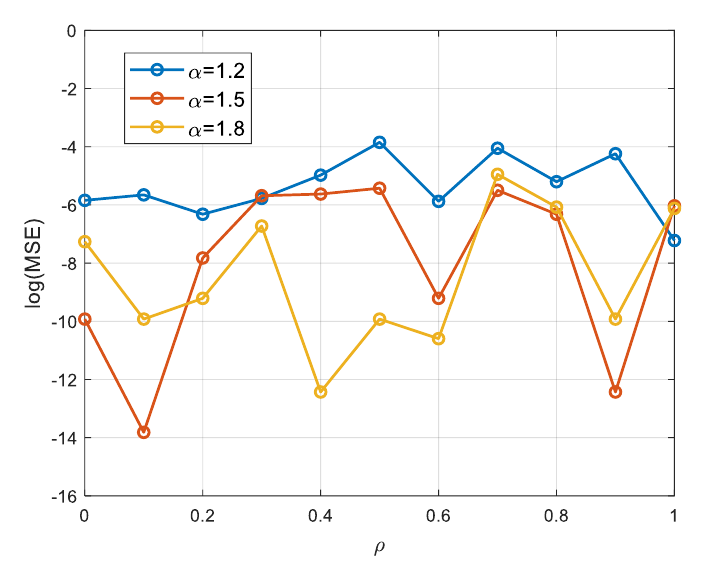}\label{fig_3_e}}
\caption{The MSE between estimation and accurate values of different parameters}
\label{fig_3}
\end{figure*}

The parameter estimation is examined to verify the performance of the proposed algorithm. Here, we only show the demodulation performance under different parameters which have a significant influence on the noise, i.e., $\alpha$ and $\rho$. Therefore, the remaining parameters are fixed and we set $\gamma_s=\gamma_g=2$, $p=5$ and the covariance matrix is \eqref{covariance_simulation_5}. The threshold $T$ in Algorithm \ref{algorithm_procedure} is set to be 0.1. We obtain the results via the mean of 50 rounds of estimation. The performance of parameter estimation is measured by the mean square error (MSE) between the estimation results and real parameters. Note that the MSE for the $\tilde{\Sigma}$ is MSE$(\tilde{\Sigma})=\frac{1}{50}\sum_{i=1}^{50}\sum_{j=2}^{p}\left[\tilde{\Sigma}(1,j)-\hat{\Sigma}(1,j)\right]^2$. We only focus on the element first row since the covariance matrix is a Toeplitz matrix.

The MSE between estimation and accurate values of various parameters are shown in Fig. \ref{fig_3}. We use the logarithmic vertical coordinate for the MSE of $\hat{\alpha}$ and $\hat{c}_1$ since they are quite small. In Fig. \ref{fig_3_a}, the MSE of $\tilde{\Sigma}$ estimation is lower when $\alpha$ is smaller since the feature of the bursty IN is more obvious. Notice that the MSE is close to 0 when $\rho=1$ because the noise degrades to WGN and $\tilde{\Sigma}$ becomes $\mathbf{I}_p$ in which case the estimation is very accurate. From Fig. \ref{fig_3_b}, there is no obvious dependence between the MSE of $\hat{\rho}$ with respect to $\alpha$ but the error is extremely low. The MSE of $\hat{\gamma}_g$ is larger when $\rho$ is close to 0 since the WGN is weak to extract the variance information. Similarly, the feature of IN clusters is vague when $\rho>0.9$ and thus, the corresponding MSE is bigger especially if $\rho=1$ in which case, however, the noise is WGN and the estimated $\gamma_s$ can be omitted here. In a word, all of the estimation accuracy is excellent with different parameter configurations.

\section{conclusion}\label{conclusion}
In this paper, we proposed a statistical model with PDF in close-form to describe the mixed Gaussian-impulsive noise with memory. The GS model admitted a weighted sum of Gaussian distribution and student distribution. Then, the parameter estimation algorithm was derived based on the empirical CF and ML estimation. Results demonstrated that our GS model could match the bursty mixed noise well and the parameter estimation was accurate in various scenarios.

\begin{appendices}
\section{Derivation of Property \ref{property_2}}\label{appendix_property_2}
\setcounter{equation}{0}
\renewcommand{\theequation}{A.\arabic{equation}}
We repeatedly indicate that $\mathbf{n}=[\mathbf{n}_1^\top,n]^\top$. Then, the conditional PDF $f_M(n|\mathbf{n}_1)$ can be expressed as \begin{align}
f_M(n|\mathbf{n}_1)=&\frac{f_M(\mathbf{n})}{f_M(\mathbf{n}_1)}=\frac{1}{{f_M(\mathbf{n}_1)}}\Bigg[\rho k_1\exp\left(-\frac{\Vert\mathbf{n}\Vert^2}{4\gamma_g^2}\right) \nonumber\\
&+\frac{(1-\rho)k_2}{\left(1+\Vert\Sigma^{-1/2}\mathbf{n}\Vert^2/\alpha\right)^\frac{\alpha+p}{2}}\Bigg].
\end{align}

According to the inverse of the partitioned matrix \cite{book5},
\begin{equation}
\Sigma^{-1}=
\left[
\begin{array}{cc}
  \mathbf{C}_1^{-1} & -\Sigma_{1,1}^{-1}\Sigma_{1,2}\mathbf{C}_2^{-1} \\
  -\mathbf{C}_2^{-1}\Sigma_{1,2}^\top\Sigma_{1,1}^{-1} & \mathbf{C}_2^{-1}
\end{array}
\right],
\end{equation}
where
\begin{align}
\mathbf{C}_1=\Sigma_{1,1}-\Sigma_{1,2}\Sigma_{2,2}^{-1}\Sigma_{1,2}^\top\in\mathbb{S}^{p-1}_{++}, \\
\mathbf{C}_2=\Sigma_{2,2}-\Sigma_{1,2}^\top\Sigma_{2,2}^{-1}\Sigma_{1,2}\in\mathbb{S}^{1}_{++}.
\end{align}

Therefore, we have
\begin{align}\label{norm_simplify}
\Vert\Sigma^{-1/2}\mathbf{n}\Vert^2=&\mathbf{n}^\top\Sigma^{-1}\mathbf{n} \nonumber\\
=&\mathbf{n}_1^\top\mathbf{C}_1^{-1}\mathbf{n}_1-x\Sigma_{1,1}^{-1}\Sigma_{1,2}\mathbf{C}_2^{-1}\nonumber\\
&-\mathbf{C}_2^{-1}\Sigma_{1,2}^\top\Sigma_{1,1}^{-1}x+\mathbf{C}_2^{-1}x^2.
\end{align}

Note that the $\mathbf{C}_2$ is a scalar and $\mathbf{C}_1^{-1}$ can be expressed by
\begin{equation}
\mathbf{C}_1^{-1}=\Sigma_{1,1}^{-1}+\Sigma_{1,1}^{-1}\Sigma_{1,2}\mathbf{C}_2^{-1}\Sigma_{1,2}^\top\Sigma_{1,1}^{-1}.
\end{equation}

Further, the \eqref{norm_simplify} becomes
\begin{align}
\Vert\Sigma^{-1/2}\mathbf{n}\Vert^2=&\mathbf{n}_1^\top\Sigma_{1,1}^{-1}\mathbf{n}_1+\frac{1}{\mathbf{C}_2}\Big[\mathbf{n}_1^\top\Sigma_{1,1}^{-1}\Sigma_{1,2}\Sigma_{1,2}^\top\Sigma_{1,1}^{-1}\mathbf{n}_1\nonumber\\
&-\mathbf{n}_1^\top\Sigma_{1,1}^{-1}\Sigma_{1,2}x-(\mathbf{n}_1^\top\Sigma_{1,1}^{-1}\Sigma_{1,2}x)^\top+x^2\Big]\nonumber\\
=&\Vert\Sigma_{1,1}^{-1/2}\mathbf{n}_1\Vert^2+\frac{(x-\mathbf{n}_1^\top\Sigma_{1,1}^{-1}\Sigma_{1,2})^2}{\mathbf{C}_2}.
\end{align}

Based on the determinant property of block matrix,
\begin{equation}
\det(\Sigma)=\det(\Sigma_{1,1})\det(\mathbf{C}_2)=\det(\Sigma_{1,1})\mathbf{C}_2.
\end{equation}

Finally, we can denote $\sigma=\det(\Sigma)/\det(\Sigma_{1,1})$ and $\delta=\mathbf{n}_1^\top\Sigma_{1,1}^{-1}\Sigma_{1,2}$ and then, the Property \ref{property_2} is proved.

\section{Derivation of property \ref{property_3}}\label{appendix_property_3}
\setcounter{equation}{0}
\renewcommand{\theequation}{B.\arabic{equation}}
According to \eqref{bursty_mixed_noise_model} and the linear property of CF, we have
\begin{align}\label{CF_weight}
\Phi_M(\mathbf{t})=&\int_{\mathbb{R}^p}f_M(\mathbf{x})\exp(-j\mathbf{t}^\top\mathbf{x})d\mathbf{x} \nonumber\\
=&\rho\Phi_G(\mathbf{t})+(1-\rho)\Phi_S(\mathbf{t})\nonumber\\
=&\rho\exp\left(-\gamma_g^2\Vert\mathbf{t}\Vert^2\right)+(1-\rho)\Phi_S(\mathbf{t})
\end{align}
where $\Phi_G(\mathbf{t})$ and $\Phi_S(\mathbf{t})$ are the CF of multivariate Gaussian distribution and student distribution, respectively. The CF of student distribution is not trivial since there is no close-formed expression. Actually,
\begin{align}\label{CF_student_mul}
\Phi_S(\mathbf{t})=&\int_{\mathbb{R}^p}\frac{k_2\exp(-j\mathbf{t}^\top\mathbf{x})}{\left(1+\Vert\Sigma^{-1/2}\mathbf{x}\Vert^2/\alpha\right)^\frac{\alpha+p}{2}}d\mathbf{x}\nonumber\\
=&\int_{\mathbb{R}^p}\frac{k_2\cos(\mathbf{t}^\top\mathbf{x})}{\left(1+\Vert\Sigma^{-1/2}\mathbf{x}\Vert^2/\alpha\right)^\frac{\alpha+p}{2}}d\mathbf{x}\nonumber\\
\end{align}

Let $\mathbf{y}=\Sigma^{-1/2}\mathbf{x}/\sqrt{\alpha}$, \eqref{CF_student_mul} becomes
\begin{align}\label{CF_simplify_1}
\Phi_S(\mathbf{t})=&\frac{k_2\sqrt{\alpha}}{\det(\Sigma^{-1/2})}\int_{\mathbb{R}^p}\cos(\sqrt{\nu}\mathbf{t}^\top\Sigma^{1/2}\mathbf{y})\nonumber\\
&\times\left(1+\mathbf{y}^\top\mathbf{y}\right)^{-\frac{\alpha+p}{2}}d\mathbf{y}. \nonumber\\
\end{align}

Combining the \eqref{CF_weight}, \eqref{CF_simplify_1} and \eqref{k_2}, we can derive the CF of the GS model is
\begin{align}
\Phi_M(\mathbf{t})=&\rho\exp\left(-\gamma_g^2\Vert\mathbf{t}\Vert^2\right)+\frac{(1-\rho)\Gamma((\alpha+p)/2)\sqrt{\alpha}}{\Gamma(\alpha/2)(\alpha\pi)^{p/2}}\nonumber\\
&\times\int_{\mathbb{R}^p}\cos(\sqrt{\nu}\mathbf{t}^\top\Sigma^{1/2}\mathbf{y})\left(1+\mathbf{y}^\top\mathbf{y}\right)^{-\frac{\alpha+p}{2}}d\mathbf{y}.
\end{align}

\section{Proof of theorem \ref{theorem_1}}\label{appendix_theorem_1}
\setcounter{equation}{0}
\renewcommand{\theequation}{C.\arabic{equation}}
The real symmetric matrix $C$ can be diagonalized as
\begin{equation}
C=P^\top\Lambda P,
\end{equation}
where $P$ is an orthogonal matrix and $\Lambda=\mathbf{diag}(\lambda_1,\lambda_2,\cdots,\lambda_p)$ is a diagonal matrix consisting of eigenvalues of $C$. Then,
\begin{equation}
\mathbf{t}_i^\top C\mathbf{t}_i=\mathbf{t}_i^\top P^\top\Lambda P\mathbf{t}.
\end{equation}

We denote that $\mathbf{s}_i=P\mathbf{t}_i$ and similarly, $\mathbf{s}_j=P\mathbf{t}_j$. To satisfy $\Vert\mathbf{t}_i\Vert\neq\Vert\mathbf{t}_j\Vert$ and $\Vert C^{1/2}\mathbf{t}_i\Vert=\Vert C^{1/2}\mathbf{t}_j\Vert$, it is equivalent to
\begin{equation}\label{equations}
\left\{
\begin{aligned}
&\sum_{k=1}^{p}[\mathbf{s}_i(k)]^2\neq\sum_{k=1}^{p}[\mathbf{s}_j(k)]^2 \\
&\sum_{k=1}^{p}\lambda_k[\mathbf{s}_i(k)]^2=\sum_{k=1}^{p}\lambda_k[\mathbf{s}_j(k)]^2
\end{aligned}
\right..
\end{equation}

Note that $C$ is not an identity matrix and $\lambda_i,i=1,\cdots,p$ are not all the same. Consequently, there are solutions for equations \eqref{equations} and the theorem \ref{theorem_1} is concluded. Then, we discuss how to choose the $\mathbf{s}_i$ and $\mathbf{s}_j$. Without loss of generality, we assume that $\lambda_1\geq\lambda_2\geq\cdots\geq\lambda_p$ and $\lambda_1\neq\lambda_2$. We set $\mathbf{s}_i(k)=\mathbf{s}_j(k),k=3,\cdots,p$ and \eqref{equations} can be simplified as
\begin{equation}\label{equations_case}
\left\{
\begin{aligned}
&[\mathbf{s}_i(1)]^2+[\mathbf{s}_i(2)]^2\neq[\mathbf{s}_j(1)]^2+[\mathbf{s}_j(2)]^2 \\
&\lambda_1[\mathbf{s}_i(1)]^2+\lambda_2[\mathbf{s}_i(2)]^2=\lambda_1[\mathbf{s}_j(1)]^2+\lambda_2[\mathbf{s}_j(2)]^2
\end{aligned}
\right..
\end{equation}

Let $\lambda_1[\mathbf{s}_i(1)]^2=\lambda_2[\mathbf{s}_j(2)]^2,\lambda_2[\mathbf{s}_i(2)]^2=\lambda_1[\mathbf{s}_j(1)]^2$ and plug them in \eqref{equations_case}, it can be calculated that $\mathbf{s}_i(1)=\sqrt{\lambda_2/\lambda_1}\mathbf{s}_j(2),\mathbf{s}_i(2)=\sqrt{\lambda_1/\lambda_2}\mathbf{s}_j(1)$ and $\mathbf{s}_j(1)/\mathbf{s}_j(2)\neq\sqrt{\lambda_2/\lambda_1}$. Then, we can choose that $\mathbf{s}_i=(\sqrt{\lambda_2/\lambda_1},\sqrt{\lambda_1/\lambda_2},1,\cdots,1)^\top,\mathbf{s}_j=(1,1,1,\cdots,1)^\top$ and $\mathbf{t}_i=P\mathbf{s}_i,\mathbf{t}_j=P\mathbf{s}_j$. Likewise, there are more $\mathbf{t}$ choices if required. Actually, the $C$ is an arbitrary positive definite matrix and all of the eigenvalues are different in practical applications.

\section{Proof of theorem \ref{theorem_2}}\label{appendix_theorem_2}
\setcounter{equation}{0}
\renewcommand{\theequation}{D.\arabic{equation}}

In order to prove the ECF $\tilde{\Phi}_M(\mathbf{t})$ converges to the CF $\Phi_M(\mathbf{t})$ in $L^2$, it is equivalent to prove that,
\begin{equation}
\lim\limits_{n\rightarrow+\infty}\mathbb{E}_{\mu_{\mathbf{X}}}\left(\left|\tilde{\Phi}_M(\mathbf{t},n)-\Phi_M(\mathbf{t})\right|^2\right)=0
\end{equation}
where
\begin{equation}\label{expression_hat_CF}
\tilde{\Phi}_M(\mathbf{t},n)=\frac{1}{n}\sum_{i=1}^{n}\exp(j\mathbf{t}^\top\mathbf{X}_i)
\end{equation}
and $\mu_{\mathbf{X}}$ represents the probability measure of the random vector $\mathbf{X}$. Then,
\begin{align}\label{CF_son}
&\mathbb{E}_{\mu_{\mathbf{X}}}\left(\left|\tilde{\Phi}_M(\mathbf{t},n)-\Phi_M(\mathbf{t})\right|^2\right)\nonumber\\
=&\mathbb{E}_{\mu_{\mathbf{X}}}\left(\tilde{\Phi}_M^2(\mathbf{t},n)\right)-2\mathbb{E}_{\mu_{\mathbf{X}}}\left(\tilde{\Phi}_M(\mathbf{t},n)\Phi_M(\mathbf{t})\right)\nonumber\\
&+\mathbb{E}_{\mu_{\mathbf{X}}}\left(\Phi_M^2(\mathbf{t})\right)
\end{align}

Note that $\Phi_M(\mathbf{t})$ is not related to $\mathbf{X}$ and therefore, we have $\mathbb{E}_{\mu_{\mathbf{X}}}\left(\Phi_M(\mathbf{t})\right)=\Phi_M(\mathbf{t})$ and \eqref{CF_son} becomes
\begin{align}\label{CF_final}
&\mathbb{E}_{\mu_{\mathbf{X}}}\left(\left|\tilde{\Phi}_M(\mathbf{t},n)-\Phi_M(\mathbf{t})\right|^2\right)\nonumber\\
=&\mathbb{E}_{\mu_{\mathbf{X}}}\left(\tilde{\Phi}_M^2(\mathbf{t},n)\right)-2\Phi_M(\mathbf{t})\mathbb{E}_{\mu_{\mathbf{X}}}\left(\tilde{\Phi}_M(\mathbf{t},n)\right)+\Phi_M^2(\mathbf{t})
\end{align}

Obviously, $\tilde{\Phi}_M(\mathbf{t},n)$ is an unbiased estimator for $\Phi_M(\mathbf{t})$ and $\mathbb{E}_{\mu_{\mathbf{X}}}\left(\tilde{\Phi}_M(\mathbf{t},n)\right)=\Phi_M(\mathbf{t})$. Based on \eqref{expression_hat_CF}, $\tilde{\Phi}_M^2(\mathbf{t},n)$ can be simplified as follows,
\begin{align}
\tilde{\Phi}_M^2(\mathbf{t},n)=\frac{1}{n^2}\Bigg[\sum_{i=1}^{n}&\exp(2j\mathbf{t}^\top\mathbf{X}_i)\nonumber\\
&+\sum_{i\neq k}\exp(j\mathbf{t}^\top(\mathbf{X}_i+\mathbf{X}_k))\Bigg]
\end{align}

Note that $\mathbf{X}_i$ and $\mathbf{X}_j$ are independent if the subscript intervals of all are larger than $M$ since the dimension of the covariance matrix is finite. We denote $I_{\alpha}$ to be the subscript set such that $I_{\alpha}^i=\{k:P(\mathbf{X}_i,\mathbf{X}_k)=P(\mathbf{X}_i)P(\mathbf{X}_k)\}$. In fact, for every fixed $\mathbf{X}_i$, the number of $\mathbf{X}_k$ which is not independent with $\mathbf{X}_i$ is finite and $\mathbf{card}((I_{\alpha}^i)^c)\leq2p-2$. Thus,
\begin{align}\label{CF_son2}
&\tilde{\Phi}_M^2(\mathbf{t},n)\nonumber\\
=&\frac{1}{n^2}\left[\sum_{i=1}^{n}\exp(2j\mathbf{t}^\top\mathbf{X}_i)+\sum_{i=1}^{n}\sum_{k\in I_{\alpha}^i}\exp(j\mathbf{t}^\top(\mathbf{X}_i+\mathbf{X}_k))\right.\nonumber\\
&\left.+\sum_{i=1}^{n}\sum_{k\in (I_{\alpha}^i)^c}\exp(j\mathbf{t}^\top(\mathbf{X}_i+\mathbf{X}_k))\right]
\end{align}

Apply the expectation on \eqref{CF_son2}, we can obtain
\begin{align}
&\mathbb{E}_{\mu_{\mathbf{X}}}\left(\tilde{\Phi}_M^2(\mathbf{t},n)\right)\nonumber\\
=&\frac{1}{n}\Phi_M(2\mathbf{t})+\left[1-\frac{\mathbf{card}((I_{\alpha}^i)^c)}{n}\right]\Phi_M^2(\mathbf{t})\nonumber\\
&+\frac{1}{n^2}\sum_{i=1}^{n}\sum_{k\in (I_{\alpha}^i)^c}\mathbb{E}_{\mu_{\mathbf{X}}}\left(\exp(j\mathbf{t}^\top(\mathbf{X}_i+\mathbf{X}_k))\right)
\end{align}

Therefore,
\begin{align}\label{CF_final_prior}
\lim\limits_{n\rightarrow+\infty}\mathbb{E}_{\mu_{\mathbf{X}}}\left(\tilde{\Phi}_M^2(\mathbf{t},n)\right)=\Phi_M^2(\mathbf{t})
\end{align}

Plug \eqref{CF_final_prior} into \eqref{CF_final}, we can conclude that $\tilde{\Phi}_M(\mathbf{t})\stackrel{L^2}{\longrightarrow}\Phi_M(\mathbf{t})$.

\section{Proof of Property \ref{property_5}}\label{appendix_property_5}
\setcounter{equation}{0}
\renewcommand{\theequation}{E.\arabic{equation}}
For the random vector $\mathbf{N}=[N_1,N_2,\cdots,N_p]^{\top}$, we can get the marginal PDF of $N_p$ by integrating the $p$-dimensional PDF, i.e.,
\begin{align}\label{marginal_pre}
f_M(n)=\int_{\mathbb{R}^{p-1}}f_M(\mathbf{n})d\mathbf{n}_1
\end{align}
where $\mathbf{n}=[\mathbf{n}_1^{\top},n]^{\top}$. Here, the $\mathbf{n}_1^{\top}$ corresponds to the random vector $\mathbf{N}_1=[N_1,N_2,\cdots,N_{p-1}]^{\top}$. As for $N_j(j<p)$, its distribution is related to the adjacent previous $(p-1)$ R.V.s and can be calculated by \eqref{marginal_pre} with the same procedure. In addition, any $p$-dimensional random vector follows the same GS model and therefore, all the component of the $[N_j,N_{j+1},\cdots,N_{j+p-1}]^{\top},j\in\mathbb{N}$ has the identical univariate distribution. Then, based on the multivariate PDF, \eqref{marginal_pre} becomes
\begin{align}\label{marginal_pre_2}
f_M(n)=&\int_{\mathbb{R}^{p-1}}\rho k_1\exp\left(-\frac{n^2+\Vert\mathbf{n}_1\Vert^2}{4\gamma_g^2}\right)d\mathbf{n}_1 \nonumber\\
&+\int_{\mathbb{R}^{p-1}}\frac{(1-\rho)k_2d\mathbf{n}_1}{\left(1+\left(\Vert\Sigma^{-1/2}\mathbf{n}\Vert^2\right)/\alpha\right)^\frac{\alpha+p}{2}}
\end{align}
and we will firstly calculate the first part in \eqref{marginal_pre_2}. Note that the Gaussian component in the GS model is isotropic. Thus,
\begin{align}\label{marginal_pre_first_part}
&\int_{\mathbb{R}^{p-1}}\rho k_1\exp\left(-\frac{n^2+\Vert\mathbf{n}_1\Vert^2}{4\gamma_g^2}\right)d\mathbf{n}_1\nonumber\\
=&\rho k_1\exp\left(-\frac{n^2}{4\gamma_g^2}\right)\int_{\mathbb{R}^{p-1}}\exp\left(-\frac{\Vert\mathbf{n}_1\Vert^2}{4\gamma_g^2}\right)d\mathbf{n}_1 \nonumber\\
=&\rho k_1\exp\left(-\frac{n^2}{4\gamma_g^2}\right)\prod_{k=1}^{p-1}\int_{\mathbb{R}}\exp\left(-\frac{z^2}{4\gamma_g^2}\right)dz\nonumber\\
=&\frac{\rho}{2\sqrt{\pi}\gamma_g}\exp\left(-\frac{n^2}{4\gamma_g^2}\right)
\end{align}

The second part in \eqref{marginal_pre_2} is more complex and the coordinate transform should be leveraged. We will obtain the marginal distribution of the second part by the derivative of its cumulative distribution function (CDF) since the $\mathbf{n}_1$ and $n$ are coupled and the relation between the $\mathbf{n}_1$ and $\mathbf{x}_1$ can not be obtained if we apply $\mathbf{w}=\Sigma^{-1/2}\mathbf{n}$. Therefore, the Jacobi matrix is impossible to derive. Based on the CDF, we have
\begin{align}
f_M(n_0)=&\frac{\partial F_M(n_0)}{\partial n_0}\nonumber\\
=&\frac{\partial[1-P(N_1>n_0)]}{\partial n_0}=-\frac{\partial P(N_1>n_0)}{\partial n_0}
\end{align}
where $F_M(N_1)$ represents the CDF of the R.V. $N_1$. Then, we utilize $\mathbf{w}=\Sigma^{-1/2}\mathbf{n}$ to simplify the integral where $\mathbf{w}=[\mathbf{w}_1^{\top},w]^{\top}$ and $\mathbf{w}_1=[w_1,w_1,\cdots,w_{p-1}]^{\top}$. Note that the $\Sigma^{-1/2}$ is the Cholesky decomposition of $\Sigma^{-1}$ and a lower triangular matrix. It can be seen that $w=\Sigma^{-1/2}(p,p)n$ and $\Sigma^{-1/2}(p,p)$ represents the element of $\Sigma^{-1/2}$ in the $p$-th row and column. For convenience, we denote it by $\Sigma^{-1/2}_p$. In this way,
\begin{align}
&P(N_1>n_0)\nonumber\\
=&\rho-F_G(n_0)+\int_{n_0}^{+\infty}\int_{\mathbb{R}^{p-1}}\frac{(1-\rho)k_2d\mathbf{n}_1dn}{\left(1+\left(\Vert\Sigma^{-1/2}\mathbf{n}\Vert^2\right)/\alpha\right)^\frac{\alpha+p}{2}}\nonumber\\
=&\rho-F_G(n_0)+\int_{n_0\Sigma^{-1/2}_p}^{+\infty}\int_{\mathbb{R}^{p-1}}\frac{\det\left(\Sigma^{1/2}\right)(1-\rho)k_2}{\left(1+\Vert\mathbf{w}\Vert^2/\alpha\right)^\frac{\alpha+p}{2}}d\mathbf{w}_1dw\nonumber\\
\end{align}
where we denote
\begin{align}
F_G(n_0)=\int_{-\infty}^{n_0}\frac{\rho}{2\sqrt{\pi}\gamma_g}\exp\left(-\frac{n^2}{4\gamma_g^2}\right)dn
\end{align}

In the next, we apply the $(p-1)$-dimensional spherical coordinate transform to simplify the integral, i.e., $w_1=r\cos\theta_1,w_j=r\prod_{k=1}^{j-1}\sin\theta_k\cos\theta_{j}(j=2,\cdots,p-2),w_{p-1}=r\prod_{k=1}^{p-1}\sin\theta_k$. The Jacobi matrix can be calculated to be $J_p=r^{p-2}\prod_{k=1}^{p-2}(\sin\theta_{k})^{p-2-k}$. Therefore,
\begin{align}
&\int_{\mathbb{R}^{p-1}}\frac{\det\left(\Sigma^{1/2}\right)(1-\rho)k_2 }{\left(1+\Vert\mathbf{w}\Vert^2/\alpha\right)^\frac{\alpha+p}{2}}d\mathbf{w}_1\nonumber\\
=&\int_{\Omega_{\Theta}}\int_{0}^{+\infty}\frac{\det\left(\Sigma^{1/2}\right)J_p(1-\rho)k_2 }{\left(1+(r^2+w^2)/\alpha\right)^\frac{\alpha+p}{2}}drd\Theta\nonumber\\
\end{align}
where $\Omega_{\Theta}$ denotes the region of $(p-2)$-fold trigonometric integral and $\Theta=(\theta_1,\cdots,\theta_{p-2})$. Before the further derivation, the following integral identity is provided,
\begin{equation}\label{integral_identity}
\int \frac{r^a}{(b+r^2)^c}=\frac{r^{a+1}}{b(a+1)}{_{2}F_1}\left(1,\frac{a+1}{c};\frac{a+c+1}{c};-\frac{r^c}{b}\right)
\end{equation}
where ${_{2}F_1}(\cdot,\cdot;\cdot;\cdot)$ denotes the Gaussian hypergeometric function and
\begin{align}\label{Gaussian_geometric_function}
{_{2}F_1}\left(a,b;c;z\right)=\sum_{k=0}^{+\infty}\frac{(a)_k(b)_k}{(c)_k}\frac{z^k}{k!},|z|\leq1
\end{align}
where $(\cdot)_k$ is the Pochhammer symbol and $(a)_k=\frac{\Gamma(a+k)}{\Gamma(a)}$. The ${_{2}F_1}\left(a,b;c;z\right)$ with $|z|>1$ can be generated by analytic extension. Then, with the aid of \eqref{integral_identity}, we have
\begin{align}\label{marginal_pre_3}
&\int_{\Omega_{\Theta}}\int_{0}^{+\infty}\frac{\det\left(\Sigma^{1/2}\right)J_p(1-\rho)k_2 }{\left(1+\left(r^2+w^2\right)/\alpha\right)^\frac{\alpha+p}{2}}drd\Theta\nonumber\\
=&(1-\rho)k_2\det\left(\Sigma^{1/2}\right)\int_{\Omega_{\Theta}}\prod_{k=1}^{p-2}(\sin\theta_{k})^{p-2-k}d\Theta\nonumber\\
&\times\int_{0}^{+\infty}\frac{r^{p-2} }{\left(1+\left(r^2+w^2\right)/\alpha\right)^\frac{\alpha+p}{2}}dr\nonumber\\
=&\int_0^{2\pi}\int_0^{\pi}\cdots\int_0^{\pi}\prod_{k=1}^{p-2}(\sin\theta_{k})^{p-2-k}d\theta_1\cdots d\theta_{p-3}d\theta_{p-2}\nonumber\\
&\times\frac{(1-\rho)k_2\det\left(\Sigma^{1/2}\right)\alpha^{\frac{\alpha+p}{2}}}{p-1}\bigg\{\bigg[r^{p-1}(\alpha+w^2)^{-\frac{\alpha+p}{2}}\nonumber\\
&\times{_{2}F_1}\left(\frac{p-1}{2},\frac{\alpha+p}{2};\frac{p+1}{2};-\frac{r^2}{\alpha+w^2}\right)\bigg]\bigg|^{r=+\infty}_{r=0}\bigg\}
\end{align}

It is rather challenging to analyze the asymptotic behavior of \eqref{marginal_pre_3} directly when $r\rightarrow+\infty$ and therefore, we introduce the following linear transformation formula,
\begin{align}\label{linear_trans}
&{_{2}F_1}\left(a,b;c;z\right)\nonumber\\
=&\frac{\Gamma(c)\Gamma(b-a)}{\Gamma(b)\Gamma(c-a)}(-z)^{-a}{_{2}F_1}\left(a,a+1-c;a+1-b;\frac{1}{z}\right)\nonumber\\
&+\frac{\Gamma(c)\Gamma(a-b)}{\Gamma(a)\Gamma(c-b)}(-z)^{-b}{_{2}F_1}\left(b,b+1-c;b+1-a;\frac{1}{z}\right)
\end{align}

Meanwhile, the Gaussian geometric function equals to 1 when $r=0$ according to the \eqref{Gaussian_geometric_function}. Substituting the parameters in \eqref{linear_trans} by $a=\frac{p-1}{2},b=\frac{\alpha+p}{2},c=\frac{p+1}{2},z=-\frac{r^2}{\alpha \tilde{n}}$, we can obtain that
\begin{align}\label{marginal_pre_4}
&\lim\limits_{r\rightarrow+\infty}\frac{r^{p-1}}{(\alpha+w^2)^{\frac{\alpha+p}{2}}}{_{2}F_1}\left(\frac{p-1}{2},\frac{\alpha+p}{2};\frac{p+1}{2};-\frac{r^2}{\alpha+w^2}\right)\nonumber\\
&=\frac{\Gamma(\frac{p+1}{2})\Gamma(\frac{\alpha+1}{2})}{\Gamma(\frac{\alpha+p}{2})}(\alpha+w^2)^{-\frac{\alpha+1}{2}}
\end{align}

In the next, the $(p-2)$-fold trigonometric integral needs to be derived. For convenience, we denote
\begin{equation}
\Upsilon(p)\triangleq\int_0^{2\pi}\int_0^{\pi}\cdots\int_0^{\pi}\prod_{k=1}^{p-2}(\sin\theta_{k})^{p-2-k}d\theta_1\cdots d\theta_{p-3}d\theta_{p-2}
\end{equation}

We have $\Upsilon(3)=2\pi$ and $\Upsilon(4)=4\pi$. $\Upsilon(2)$ is derived at the end as the special case. Based on the Wallis formula, the $\Upsilon(p)$ can be obtained that
\begin{equation}
\Upsilon(p)=\left\{
\begin{aligned}
&2\pi,\hspace{4cm} p=3\\
&4\pi,\hspace{4cm} p=4\\
&2^{\frac{8-p}{2}}\pi^{\frac{p-2}{2}}\prod_{k=1}^{\frac{p-4}{2}}\frac{1}{2k+1},\hspace{1.1cm} \text{ $p-3$ is odd}, p\geq5 \\
&2^{\frac{7-p}{2}}\pi^{\frac{p-1}{2}}\frac{p-4!!}{p-3!!}\prod_{k=1}^{\frac{p-5}{2}}\frac{1}{2k+1}, \text{ $p-3$ is even}, p\geq5.
\end{aligned}
\right.
\end{equation}

Combining with \eqref{marginal_pre_3} and \eqref{marginal_pre_4}, the marginal distribution of the GS model is
\begin{align}\label{marginal_1}
&f_M(n)\nonumber\\
=&\frac{\partial F_G(n)}{\partial n}-\frac{\partial }{\partial n}\int_{n\Sigma^{-1/2}_p}^{+\infty}\int_{\mathbb{R}^{p-1}}\frac{\det\left(\Sigma^{1/2}\right)(1-\rho)k_2}{\left(1+\Vert\mathbf{w}\Vert^2/\alpha\right)^\frac{\alpha+p}{2}}d\mathbf{w}_1dw\nonumber\\
=&\frac{\rho}{2\sqrt{\pi}\gamma_g}\exp\left(-\frac{n^2}{4\gamma_g^2}\right)+(1-\rho)k_m(\alpha+(n\Sigma^{-1/2}_p)^2)^{-\frac{\alpha+1}{2}}
\end{align}
where
\begin{align}\label{k_m}
k_m=\frac{\Upsilon(p)\alpha^{\frac{\alpha}{2}}\Gamma(\frac{p+1}{2})\Gamma(\frac{\alpha+1}{2})}{(p-1)\Gamma(\frac{\alpha}{2})\pi^{\frac{p}{2}}}
\end{align}

In fact, the normalization factor $k_m$ can be simplified by the property of the PDF, i.e.,
\begin{align}
\int_{-\infty}^{+\infty}k_m(\alpha+(n\Sigma^{-1/2}_p)^2)^{-\frac{\alpha+1}{2}}dn=1
\end{align}
and therefore,
\begin{align}\label{new_k_m}
k_m=\frac{\alpha^{\frac{\alpha}{2}}\Sigma^{-1/2}_p\Gamma(\frac{\alpha+1}{2})}{2\Gamma(\frac{3}{2})\Gamma(\frac{\alpha}{2})}
\end{align}

Finally, we figure out the $\Upsilon(2)$ as a special case. In this case, there is no need to apply coordinate transform. Here, we will omit the detailed derivation to avoid repetition and similar procedures are based on \eqref{marginal_pre_first_part} and \eqref{marginal_pre_3}. When $p=2$,
\begin{align}\label{marginal_p_2}
f_M(n)=&\frac{\rho}{2\sqrt{\pi}\gamma_g}\exp\left(-\frac{n^2}{4\gamma_g^2}\right)\nonumber\\
&+2\int_{0}^{+\infty}\frac{(1-\rho)k_2}{\left(1+\left(z^2+(n-\delta)^2/\sigma\right)/\alpha\right)^\frac{\alpha+2}{2}}dz\nonumber\\
=&\frac{\rho}{2\sqrt{\pi}\gamma_g}\exp\left(-\frac{n^2}{4\gamma_g^2}\right)+2(1-\rho)k_2\sqrt{\alpha}\nonumber\\
&\times\frac{\Gamma(\frac{3}{2})\Gamma(\frac{\alpha+1}{2})}{\Gamma(\frac{\alpha+2}{2})}\left(1+\frac{n^2}{\alpha}\right)^{-\frac{\alpha+1}{2}}
\end{align}

Compared the \eqref{marginal_1} with \eqref{marginal_p_2}, we can conclude that $\Upsilon(2)=2$. In fact, based on \eqref{k_m} and \eqref{new_k_m}, we can obtain a simpler expression for $\Upsilon(p)$ as follows,
\begin{align}\label{new_Upsilon}
\Upsilon(p)=\frac{(p-1)\pi^{\frac{p}{2}}}{2\Gamma(\frac{3}{2})\Gamma(\frac{p+1}{2})}
\end{align}

\section{Proof of Property \ref{property_6}}\label{appendix_property_6}
\setcounter{equation}{0}
\renewcommand{\theequation}{F.\arabic{equation}}
The $r$-th moment of $N_1$ is expressed as follows,
\begin{align}
&\mathbb{E}_{\mu_{N_1}}[|N_1|^r]\nonumber\\
=&\int_{-\infty}^{+\infty}|n|^rf_M(n)dn\nonumber\\
=&\frac{\rho}{\sqrt{\pi}\gamma_g}\int_{0}^{+\infty}n^r\exp\left(-\frac{n^2}{4\gamma_g^2}\right)dn+\frac{\alpha^{\frac{\alpha}{2}}\Sigma^{-1/2}_p\Gamma(\frac{\alpha+1}{2})}{\Gamma(\frac{3}{2})\Gamma(\frac{\alpha}{2})}\nonumber\\
&\times(1-\rho)\int_{0}^{+\infty}|n|^r(\alpha+(n\Sigma^{-1/2}_p)^2)^{-\frac{\alpha+1}{2}}dn\nonumber\\
=&\frac{(2\gamma_g)^{r}\rho}{\sqrt{\pi}}\Gamma\left(\frac{r+1}{2}\right)+\frac{(1-\rho)\alpha^{\frac{\alpha}{2}}\Sigma^{-1/2}_p\Gamma(\frac{\alpha+1}{2})}{\Gamma(\frac{3}{2})\Gamma(\frac{\alpha}{2})}\int_{0}^{+\infty}n^r\nonumber\\
&\times(\alpha+(n\Sigma^{-1/2}_p)^2)^{-\frac{\alpha+1}{2}}dn\nonumber\\
=&\frac{(2\gamma_g)^{r}\rho}{\sqrt{\pi}}\Gamma\left(\frac{r+1}{2}\right)+\frac{(1-\rho)\alpha^{\frac{r}{2}}\Gamma(\frac{r+3}{2})\Gamma(\frac{a-r}{2})\left(\Sigma^{-1/2}_p\right)^{r}}{\Gamma(\frac{3}{2})\Gamma(\frac{\alpha}{2})(r+1)}
\end{align}

The last equation can be obtained based on \eqref{integral_identity}, \eqref{linear_trans} and $r<\alpha$.

\end{appendices}

\footnotesize
\bibliographystyle{IEEEtran}
\bibliography{ref}

\end{document}